\documentclass[12pt]{article}
\usepackage[top=1in, bottom=1in, left=1in, right=1in]{geometry}
\usepackage[english]{babel}
\usepackage{atbegshi,cite}
\usepackage{amsmath, amssymb, amsbsy, amstext, amsthm, array, booktabs, IEEEtrantools, graphicx, cite, braket, float, tikz, subcaption, quantikz, setspace, listings, authblk, multirow, simplewick, textgreek,upgreek,cancel,braket,bigints}
\usepackage{bm} 
\usepackage{graphicx}
\usepackage{amsfonts}
\usepackage[titletoc]{appendix}
\usepackage{setspace}
\usepackage{comment}
\usepackage{tensor} 
\usepackage{xcolor}
\usepackage{subcaption}
\usepackage[labelfont=bf, font=footnotesize]{caption}
\usepackage[inline]{enumitem}
\usepackage[usenames,dvipsnames,table]{xcolor}
\usepackage[colorlinks=true,urlcolor=blue
,anchorcolor=blue,citecolor=blue,filecolor=blue,linkcolor=blue,menucolor=blue
]{hyperref}
\usepackage{algorithm}
\usepackage{algpseudocode}
\usepackage{adjustbox}

\usepackage{fancyhdr}
\usepackage[yyyymmdd,hhmmss]{datetime}
\renewcommand{\mathbf}{\bm} 

\graphicspath{ {./} {../} }
\begin{document}

\newcommand{\authors}{}
\newcommand{\titleLong}{Time-Domain Observables for Resonant Scattering in Quantum Simulations of Lattice Gauge Theory}

\title{\huge \titleLong{}}
\author[1]{Praveen Balaji}
\author[1]{Patrick Draper}
\author[2]{Enrico Rinaldi}
\affil[1]{Department of Physics, University of Illinois, Urbana, IL 61801}
\affil[2]{Quantinuum, Partnership House, London, UK}

\date{}

\fancypagestyle{firstPage}{
  \fancyhead[L]{\small \textbf{\authors{}}}
  \fancyfoot{}
}

\maketitle

\begin{abstract}
    Quantum computing may allow unprecedented access to nonequilibrium quantum dynamics, including problems of interest in particle physics, like the real-time development of scattering processes.
    Although fault-tolerant simulations at scale are still in the future, recent advancements toward early fault-tolerant simulations motivate building increasingly elaborate benchmark problems in order to assess state-of-the-art capabilities and test new algorithms and observables. 
    In this work, we study glueball scattering on a line-of-plaquettes lattice in pure $SU(3)$ lattice gauge theory. We characterize scattering of the lowest-energy glueball states using classical exact diagonalization methods, identifying the continuum resonance structure using real-time simulations and approximate analytic methods, and we propose time-domain observables for extracting properties of the resonance from quantum simulations. We furthermore construct a toy model that shares the spectral features of the lattice gauge theory and demonstrate extraction of the observables. Feasibility on near-term hardware is demonstrated through simulation of a resonant collision of this toy model on Quantinuum System Model \texttt{H2-2}, with a 14-plaquette chain encoded in 28 qubits. The Trotterized circuit after optimization and compression requires $\sim 25n$ entangling gates for $n$ Trotter steps.
\end{abstract}
\newpage 
\tableofcontents
\section{Introduction and Background}
\indent With the advent of quantum computers that contain hundreds of high-fidelity qubits, full simulation of non-perturbative real-time dynamics in quantum field theories (QFTs) is within reach \cite{Jordan2012,Bauer2023Roadmap}. A family of QFTs that are of great interest to simulate are non-Abelian gauge theories, with the biggest target being quantum chromodynamics (QCD). Such theories have seen a large body of work in encoding, state preparation, and Hamiltonian evolution on a quantum computer \cite{Atas2023QCD, Farrell2023AxialGauge, Kadam2025LSH, Kadam2026LSHOperators, Ciavarella2021SU3, Ciavarella2022Vacuum, Byrnes2006LGT, Kan2021QCDQED, Illa2025Honeycomb, Klco2020SU2, Cataldi2025Strings, Carena2022Improved, Modi2026LargeNc, Bergner2026Minimal, Gupta2026StringBreaking, k8f6-yft8, Balaji_2026, OsorioPerez2025PrimitiveGates, Hidalgo2026LatticeQCD, PhysRevD.105.114508, PhysRevD.111.114516, Zohar2015Formulation}. Simulations can illuminate processes such as confinement and hadronization, as well as serve as testbeds for various scattering experiments with hadrons and gluons \cite{Bauer2023Fundamental, DiMeglio2024HEP, Halimeh2025Dynamics}. Scattering is a particularly rich use-case where a breadth of observables of the theory that can inform experiment (resonances, cross-sections, transient states) can be extracted from a single collision. It also provides a demanding setting for current quantum simulation techniques, since it requires state preparation of spatially separated particles, real-time evolution of their trajectories in the interaction region, and readout of the collision products. Quantum simulations of particle scattering have received considerable attention over the years \cite{davoudi2025,farrell2025,Schuhmacher2025Scattering, Zemlevskiy2025Scattering, Surace2021MesonScattering, Chai2026MesonScattering, Chai2026ResourceEfficient, Joshi2025Qudit}. Related developments include interacting wave-packet preparation on quantum computers \cite{Davoudi2024WavePackets, 2nb4-171l, Turco2024BoundStateScattering, Surace2026StatePreparation}, tensor-network studies of real-time meson scattering \cite{Rigobello2021Scattering, Bennewitz2025MesonScattering, Belyansky2024Collisions}, and quantum algorithms for extracting scattering phase shifts along with other real-time scattering observables \cite{Gustafson2021TimeDelay,Sharma2024PhaseShifts, Briceno2026RESO, Burbano2026, Li2024ScatteringAmplitude, Ciavarella2020ParticleDecays}. However, there have not been many attempts to tackle the complexity of scattering in non-Abelian gauge theories. A first step is to estimate the resources required to simulate a QCD collision on resonance and to extract its quantitative features on a quantum device. 

As a step toward real-time simulation of lattice QCD, we study low-energy scattering in $SU(3)$ lattice gauge theory in the intermediate- to strong-coupling regime. We consider one-dimensional collisions on a chain of plaquettes with periodic boundary conditions (PBCs), where the colliding particles are the lowest glueball excitations\footnote{\emph{Glueballs} is the technical term used to refer to composite bound states in a pure gauge theory.}. Through classical simulation, we quantify a resonance with a localized heavier glueball state. The structure of the resonance is mapped onto the Friedrichs model \cite{Friedrichs1948}, where a discrete state couples to an approximate continuum of scattering states. Near individual crossings, it is observed, through spectra and Rabi oscillation of the resonant scattering mode, that the dynamics can be approximately described by an effective two-level system consisting of the localized state and the resonant scattering mode. The avoided crossing gap ($\Delta_{\text{AC}}$) approximates the coupling to the resonant scattering state and, together with the continuum density of states, provides a connection to the resonance width in the infinite-volume theory. While the extraction of resonance properties from finite-volume spectra is a well-established program in lattice QCD \cite{Briceno2018Resonances, DeGrand2006LatticeMethods, Luscher1990VariationalMethod, Blossier2009GEVP, Rummukainen1995MovingFrame, Luscher1991Scattering, Dudek2013RhoResonance, Wilson2015CoupledChannel, Dudek2016a0Resonance, Briceno2018ScatteringReview}, here we investigate a spectral description of the resonance together with observables accessible through real-time evolution. 

To understand what is required to observe such a resonance on quantum hardware, we construct collisions and decays for a minimal effective model containing the vacuum, a propagating low-energy glueball state, and a higher-energy localized state representing the resonant degree of freedom. We construct circuits for wave-packet and scattering state preparation, and optimize the Trotterized real-time evolution for two-qubit gate count and circuit depth. We use these circuits to study both wave-packet collisions and the coherent decay of an isolated scattering state. The latter provides a direct measurement of the avoided crossing gap through the oscillation period. We  implement the optimized collision circuits on Quantinuum System Model H2~\cite{Moses2023H2}, demonstrating a pathway from the characterization of a resonance in a non-Abelian lattice gauge theory to a reduced quantum simulation of its essential real-time dynamics.
\subsection{Kogut-Susskind Hamiltonian in the Reduced Electric Basis}
\begin{figure}[H]
\begin{center}
\includegraphics[width=1.00\textwidth]{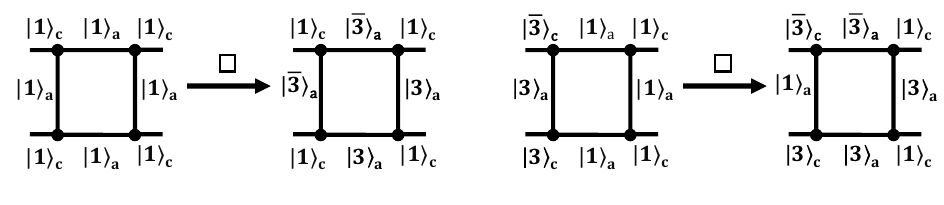} %
\caption{ Examples of the action of the $\Box$ operator on a plaquette. The subscripts ``a'' and ``c'' denote the active and control links respectively. The singlet degrees of freedom (DoFs) on the vertices have been omitted for clarity. As seen in the picture on the left, the $\Box$ operator generates flux: the vacuum $\ket{1}$ state transforms to a plaquette of all $\ket{3}$ states, which has non-zero electric density. The picture on the right demonstrates the change of flux as well, but with a different transition. Only one of the basis states the $\Box$ operator generates upon acting on the initial state is shown.} 
    \label{fig:box_action}
\vspace*{-0.5cm}
\end{center}
\end{figure}

We work with the $SU(3)$ Kogut-Susskind Hamiltonian \cite{KogutSusskind1975} where the only degrees of freedom are the link variables, defined as 
\begin{equation}
\begin{aligned}
H = \frac{g^{2}}{2a}\sum_{x, \mu} \sum_{l} \left(E^{l}(x, \mu)\right)^{2} + \frac{1}{g^{2}a} \sum_{P \in {\text{plaquettes}}} \left(2N - \left(\Box_{P} + \Box^{\dagger}_{P}\right)\right)
\end{aligned}
\end{equation}
for coupling constant $g$ and lattice spacing $a$. The constant $N$ is a conventional shift that will be ignored below and we take units $a=1$. The magnetic term of the Hamiltonian contains the plaquette operator $\Box$, which is defined with link operators $U_{\textbf{3}}$ in the fundamental representation as 
\begin{equation}
\begin{aligned}
\Box = \text{Tr}\left(U_{3}(x, \mu)U_{3}(x + \mu, \nu)U_{3}^{\dagger}\left(x + \nu, \mu \right)U_{3}^{\dagger}\left(x, \nu \right) \right).
\end{aligned}
\end{equation}

We work in the electric basis (eigenbasis of $E^{2}$). The states are labeled by an irrep and associated quantum numbers (total isospin $T$, z-isospin $T_{z}$, and hypercharge $Y$). Since $SU(3)$ is non-Abelian, the link Hilbert space is a tensor product of a left space and right space that undergo left and right action under the Hamiltonian, respectively. The states are of the form $\ket{R,  Q_{L}} \otimes \ket{R,  Q_{R}}$ for an irrep $R$ and quantum number sets ${Q_L, Q_R}$. As discussed in Refs.~\cite{Ciavarella2021SU3,k8f6-yft8}, the electric basis can be reduced by projecting onto the singlet states at every vertex. The electric basis $\ket{\omega}$ for links around a site $s$ contains elements of the form 
\begin{equation}
\begin{aligned}
\ket{\omega}_{s} \supset \sum_{Q_{LR}} \braket{1, \Gamma_{s}| \otimes_{\ell(s)} R, Q_{L, R}} \ket{\otimes_{\ell(s)} R, Q_{L, R}} 
\end{aligned}
\end{equation}
where the internal quantum numbers are summed over to form the singlet and the $\Gamma_{s}$ index is a multiplicity index, since there may be multiple distinct singlets that the links arriving at a site may be projected onto. The end result of this projection is the reduced electric basis, where the global electric basis state is $\ket{\omega} \in \{\bigotimes_{s \in S, \ell(s)} \ket{\Gamma_{s}} \ket{R_{\ell}}\}$; each link $\ell$ just contains an irrep index $R_{\ell}$ and each vertex contains a multiplicity index $\Gamma_{s}$. The $\Box$ matrix elements, which operate on plaquettes, also change accordingly with the additional Clebsch-Gordan (CG) coefficients from the projection. The calculation of the $\Box$ matrix elements is given in \cite{k8f6-yft8}. Fig.~\ref{fig:box_action} shows what a plaquette state and transition look like in the reduced electric basis.
\subsection{Simulation Scheme}

\begin{figure}[H]
    \centering

    \begin{minipage}[c]{0.55\textwidth}
        \centering
        \includegraphics[width=\linewidth]
        {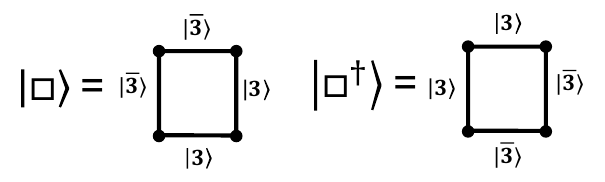}
    \end{minipage}
    \hspace{0.7cm}
    \begin{minipage}[c]{0.25\textwidth}
        \centering
        \includegraphics[width=\linewidth]
        {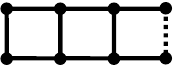}
    \end{minipage}

    \caption{\textbf{Left:} The lowest electric energy glueball states at strong coupling. We consider the scattering of the particles formed from these excitations and their dressings. \textbf{Right:} An example of a chain of plaquettes with PBCs. }
    \label{fig:box_3x}
\end{figure}

A standard link-level truncation is to restrict the Hilbert space to low-lying electric Casimirs. For $g \gtrsim 1$ this is also a low-energy projection. In addition to truncating on the link irreps a more granular truncation is introduced by truncating on the electric energy density at a site (the ``$B$'' truncation scheme \cite{Balaji_2026}). We combine the two truncations, using $B=6.0$ and truncating the link local Hilbert space to $\{\mathbf{1}, \mathbf{3}, \mathbf{\overline{3}}, \mathbf{8}\}$ (excluding $\mathbf{6}, \mathbf{\overline{6}}$). This truncation allows the $\mathbf{8}$ to appear in two different kinds of vertex configurations --- $\{\mathbf{3}, \mathbf{\overline{3}}, \mathbf{8}\}$ and $\{\mathbf{1}, \mathbf{8}, \mathbf{8}\}$ --- but excludes $\{\mathbf{8}, \mathbf{8}, \mathbf{8}\}$. At strong coupling, the lowest-energy glueball states are single $\Box$ and $\Box^{\dagger}$ excitations from the vacuum, which are plaquette states with all $\mathbf{3}$ and $\mathbf{\overline{3}}$ irreps on the links, as depicted in Fig.~\ref{fig:box_3x}. For shorthand, we will refer to the particles formed from superpositions of these excitations as $\Box_{3}$, even though these excitations will mix with higher-energy states as the coupling is decreased. 

We consider scattering on one-dimensional plaquette chains (``$d=3/2$'' for dimension $d$) with PBCs, also depicted in Fig.~\ref{fig:box_3x}. To create colliding particles that are well-separated and localized around particular momenta, the number of plaquettes $N$ must be sufficiently large. A larger $N$ comes at a simulation cost. For our classical exact diagonalization simulations, we look at the plaquette chains with $N = 12$ and $N = 16$, both considered with additional truncations on the global energy density. Details of the classical simulation techniques used are given in App.~\ref{app:global}.

\section{Classical Characterization of $\Box_{3}$ Scattering}
\subsection{Single-particle Construction}
\begin{figure}[H]
\centering
\hspace{-1cm}
\includegraphics[width=0.50\textwidth]{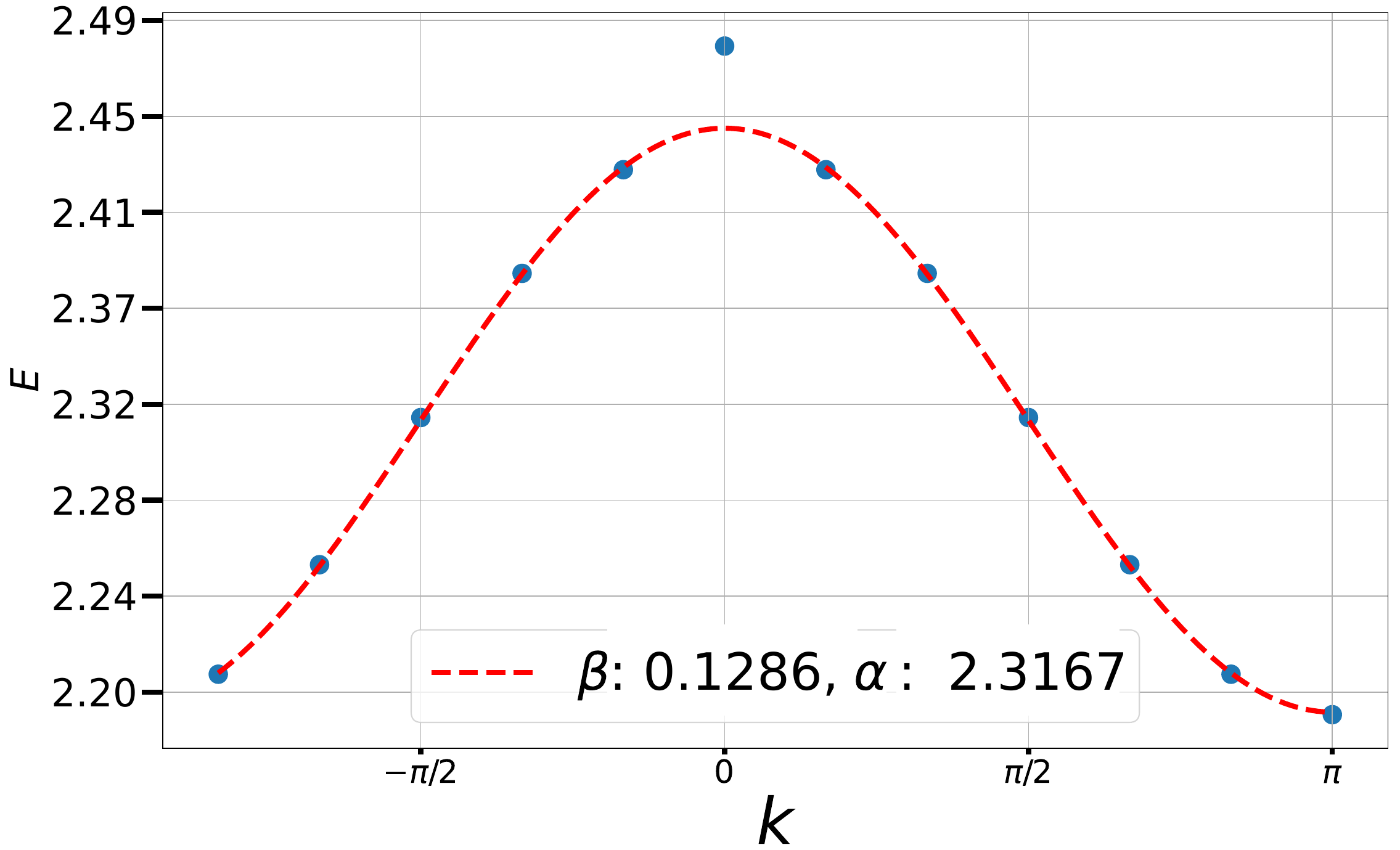} %
\hspace{-0.1cm}
\includegraphics[width=0.50\textwidth]{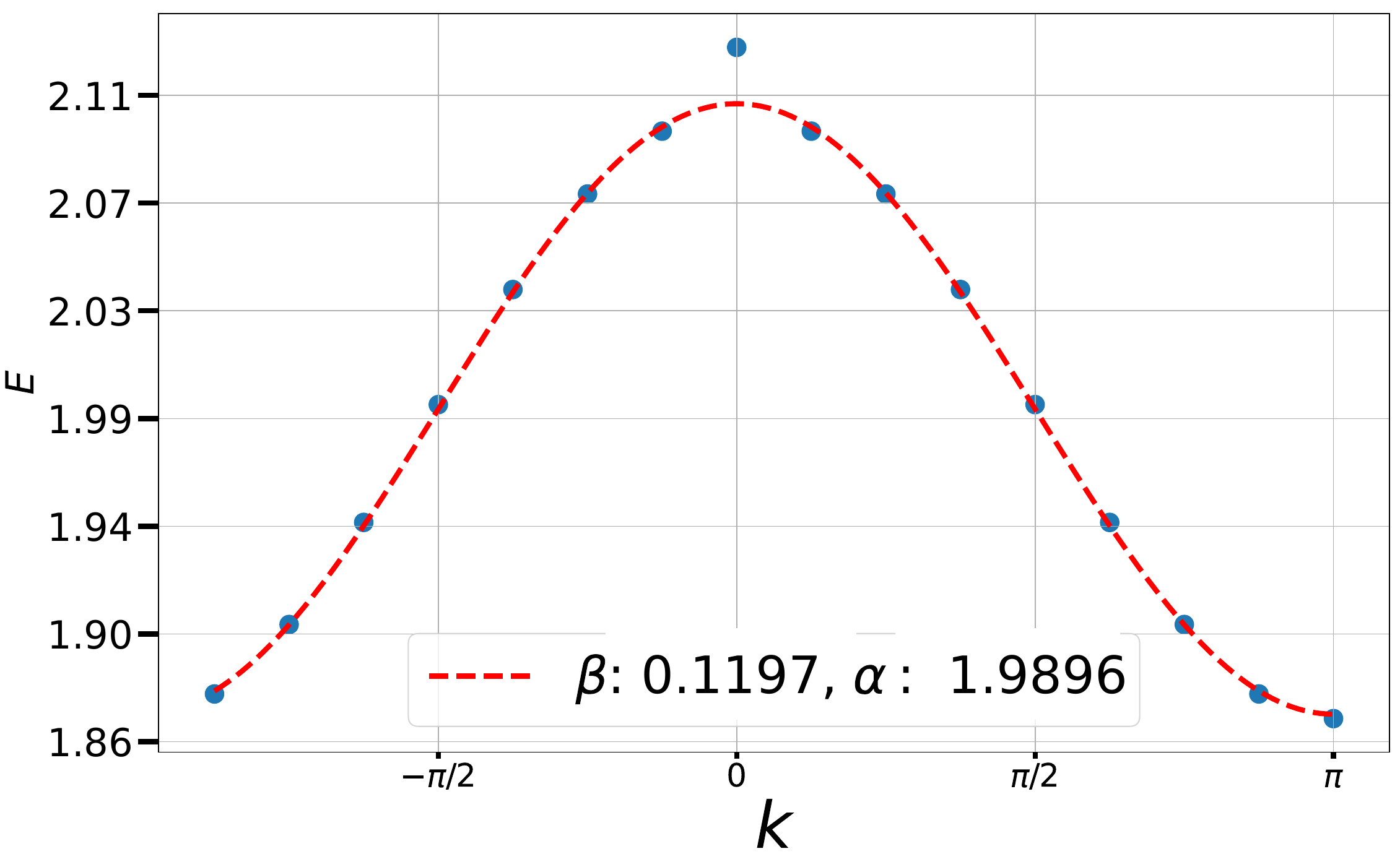} 
\caption{Single-particle dispersions for $N=12$ (\textbf{left}) and $N=16$ (\textbf{right}) for $g=1.3$. The departure of the $k=0$ state from the fitted dispersion is a truncation effect.}
\label{fig:single_particle_fits}
\end{figure}

To create a particle with momentum $k_{0}$ located at plaquette $n_{0}$, we construct a Gaussian wave-packet over the momentum modes, 
\begin{equation}
\begin{aligned}
\ket{\psi(k_0, n_{0})} = \frac{1}{\mathcal{N}}\sum_{k} e^{i  k n_{0}}e^{-(k-k_0)^{2}/2\sigma_{k}^{2}}\ket{k}
\label{eqn:wavepacket}
\end{aligned}
\end{equation}
where the momentum modes are plane waves over superpositions of the $\Box_{3}$ particles. To find these modes and their dispersion, we use strong-coupling perturbation theory to determine how the plaquette operator splits the degenerate spectrum. 
By rescaling the Hamiltonian, we can bring it into a canonical form for perturbation theory 
\begin{equation}
\begin{aligned}
H = \sum_{x, \mu} E^{2}(x, \mu) - \lambda \sum_{P} \left(\Box_{P} + \Box_{P}^{\dagger}\right)
\end{aligned}
\end{equation}
where $\lambda = 2 / g^{4}$. At first order, $\{\ket{\Box}_{n}\} \leftrightarrow \{\ket{\Box^{\dagger}}_{n}\}$ on the same plaquette, which is a $\sigma_{x}$-like Hamiltonian. Thus, the resulting eigenvalues are $\mp$ 1 with eigenstates $\left(\frac{\ket{\Box_{n}} + \ket{\Box^{\dagger}_{n}}}{\sqrt{2}}\right) = \ket{+}_{n}$ and $\left(\frac{\ket{\Box}_{n} - \ket{\Box^{\dagger}}_{n}}{\sqrt{2}}\right) = \ket{-}_{n}$, respectively. The momentum modes can be built on either excitation; for the rest of the paper we focus on the $\ket{+}$ states. The momentum modes are superpositions of $\ket{+}$ states that are also eigenstates of the lattice translation operator. The $N$ states over $N$ plaquettes explicitly are
\begin{equation}
\begin{aligned}
\ket{k} = \frac{1}{\sqrt{N}} \sum_{n=0}^{N-1} e^{- \frac{2 \pi i n_{k} n}{N}} \left(\frac{\ket{\Box}_{n} + \ket{\Box^{\dagger}}_{n}}{\sqrt{2}}\right) = \frac{1}{\sqrt{N}} \sum_{n=0}^{N-1} e^{ -\frac{2 \pi i n_{k} n}{N}} \ket{+}_{n}
\end{aligned}
\end{equation}
where $n_{k}$ is the crystal momentum $\in [N]$ associated with physical momentum ($ k = 2 \pi n_{k} / N$). With $k \in [-\pi, \pi]$ and $\omega(k) = \omega(-k)$, the dispersion relation is of the form 
\begin{equation}
\begin{aligned}
\label{eqn:dispersion_form}
\omega(k) = \alpha + \sum_{n=1}^{\lfloor N/2 \rfloor}
\beta_{n} \cos(n k)
\end{aligned}
\end{equation}
The $\beta_{n}$ are $O(\lambda^{2})$ and thus, in the intermediate- to strong-coupling regime, $\beta_{n} \ll \alpha$; the particles are massive and the spectrum  is quite flat. App.~\ref{app:effective} gives the second-order ($\lambda^{2}$) calculation of these coefficients for the irrep and $B$ truncation, where the effective potential, a single-plaquette hopping term, yields a cosine dispersion. Fig.~\ref{fig:single_particle_fits} shows the numerical dispersion in the intermediate coupling regime ($g \leq 1.7$ with $\lambda^{3} \sim 10^{-2}$). There is an ``anomaly" for the energy of the $k=0$ state, which does not fit in the cosine dispersion. This anomaly is due to the truncation, which allows the raw vacuum to mix more strongly with the dressed $k=0$ state and push its energy up.

Even in the intermediate regime where higher orders of $\lambda$ have an effect, the dispersion is still effectively captured by $\alpha + \beta \cos(k)$. This occurs because $n$-th order hopping terms are additionally suppressed by the energy density itself in the form $(1 / \alpha)^{n}$; in order for the single plaquette excitation to hop, it mixes with either the vacuum or a two plaquette excitation that have $\sim \Delta \alpha$ electric energy from the single-plaquette excitation.

\begin{figure}[H]
\centering

\begin{minipage}{\textwidth}
    \centering
    \includegraphics[width=0.48\textwidth]{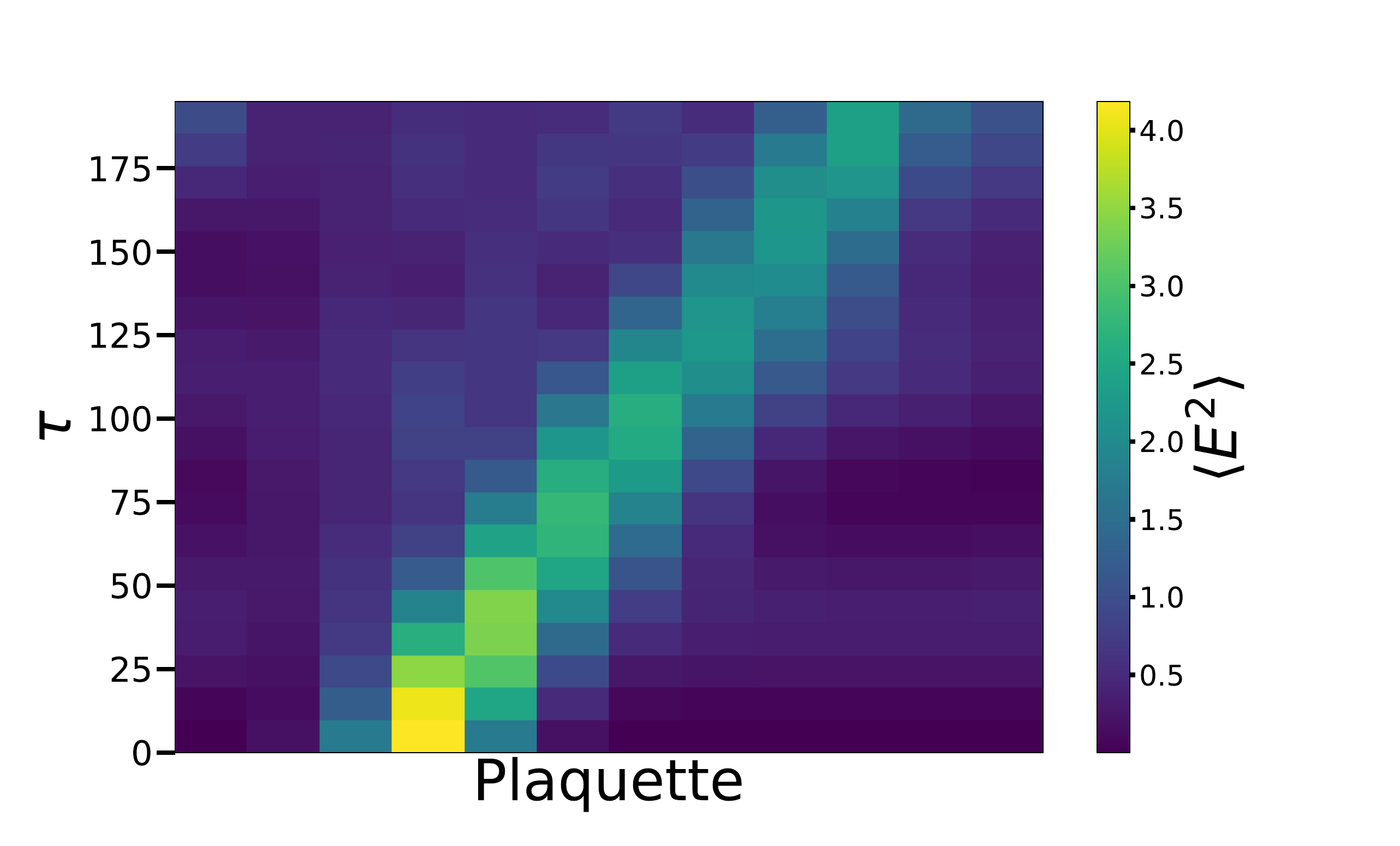}
    \hspace{0.02\textwidth}
    \includegraphics[width=0.48\textwidth]{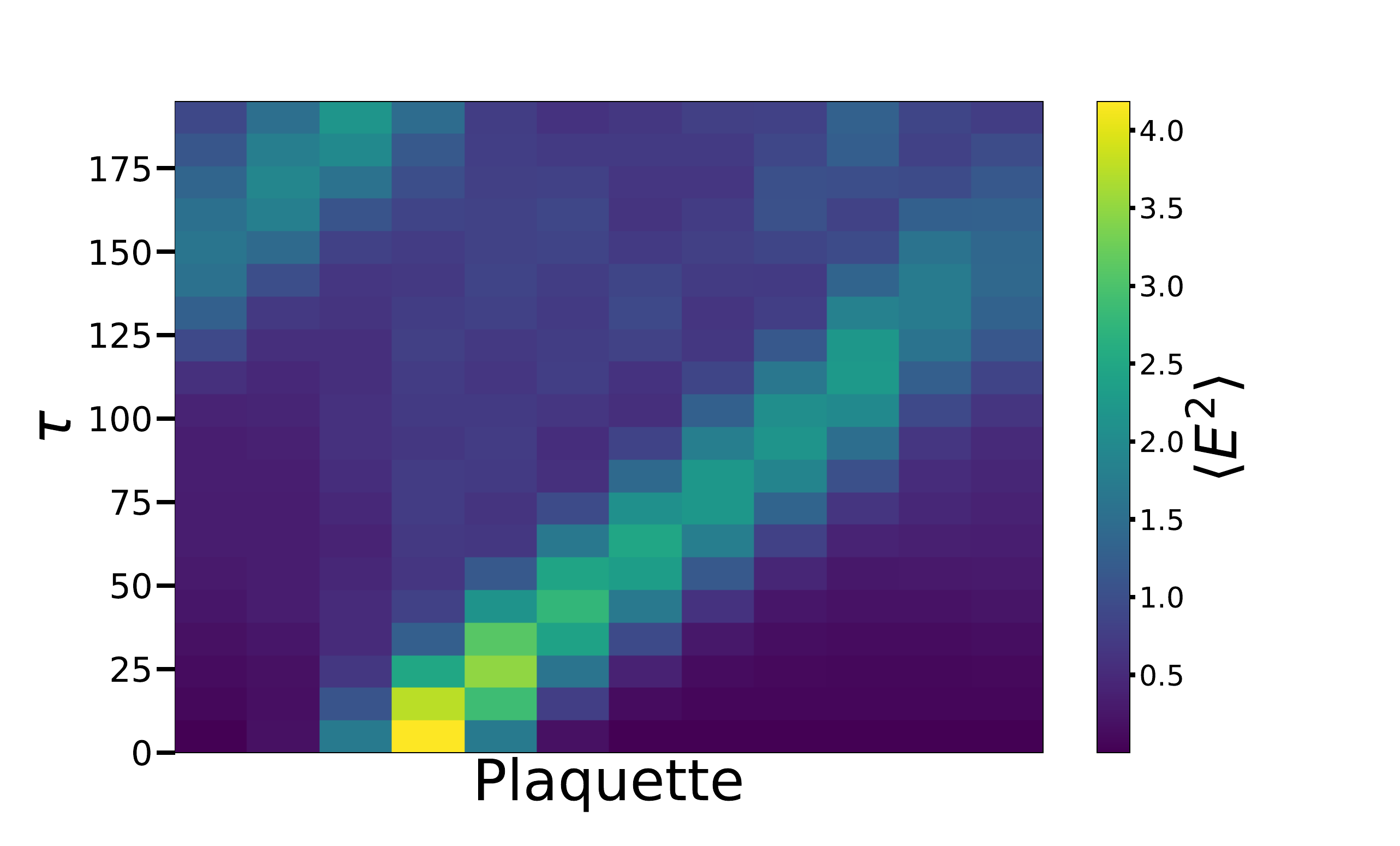}
\end{minipage}

\caption{Evolution of a single packet over 12 plaquettes for
$g=1.5$ and $g=1.4$, both with $\sigma_k=2.5$ and $k_0=3$.
The packet is tracked by electric energy density across the chain.
}
\label{fig:general_scattering}

\end{figure}

With the momentum modes and the wave-packet of Eq.~(\ref{eqn:wavepacket}), localized $\Box_{3}$ particles with definite momentum $k_{0}$ can be created at any coupling by first creating the known eigenstate at strong-coupling and then by doing an adiabatic ramp down. Fig.~\ref{fig:general_scattering} shows packet evolution tracked by electric energy density per plaquette for two different couplings $g$, where the significance is that packet group velocity increases with smaller $g$. 

\subsection{Characterization of Collisions}
With the single-particle construction, we can proceed to two-particle scattering and characterize the interaction.

\subsubsection{Product State Construction}
\begin{figure}[t]
\centering
\hfill 
\hskip -1.5em
\includegraphics[width=0.50\textwidth]{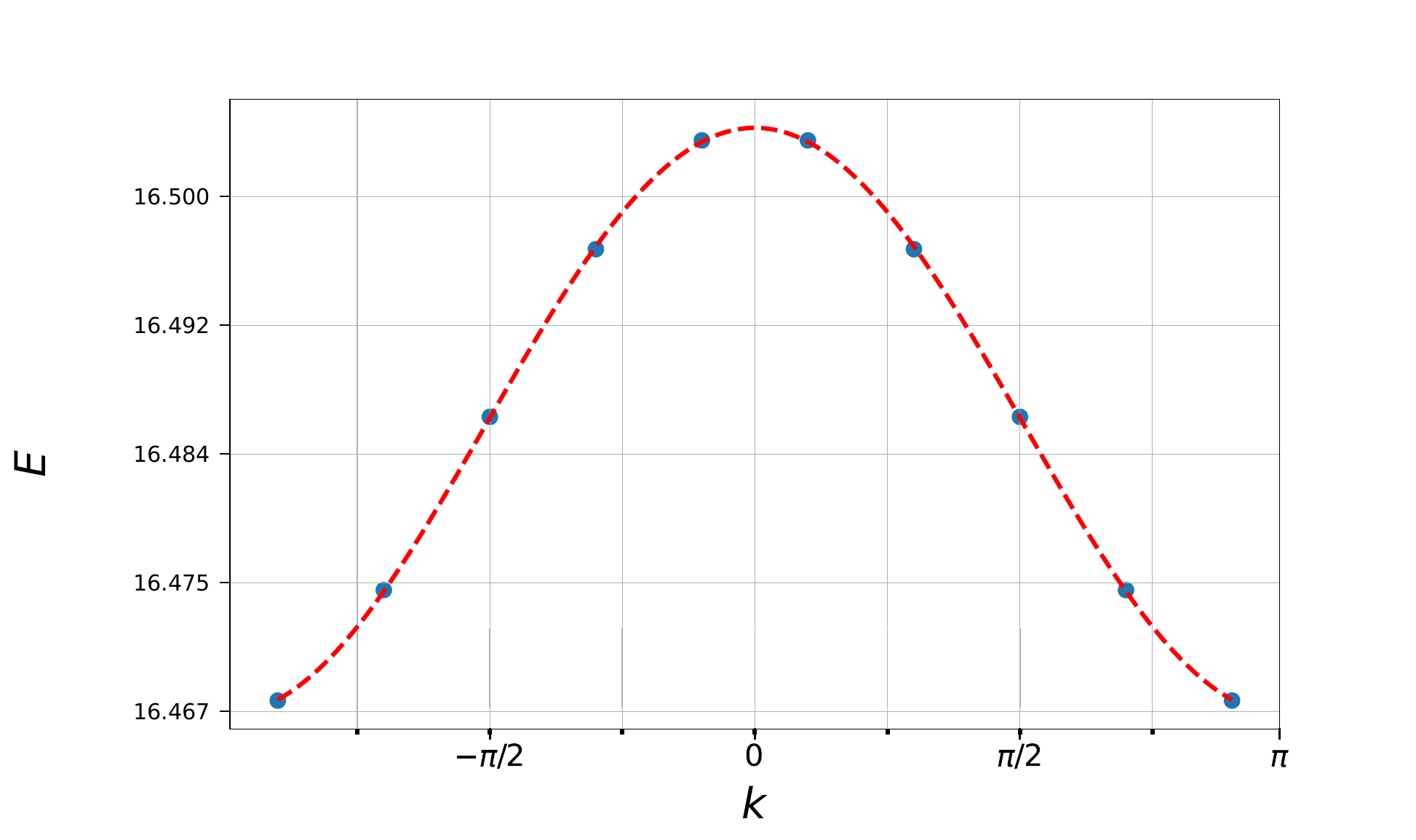} %
\includegraphics[width=0.50\textwidth]{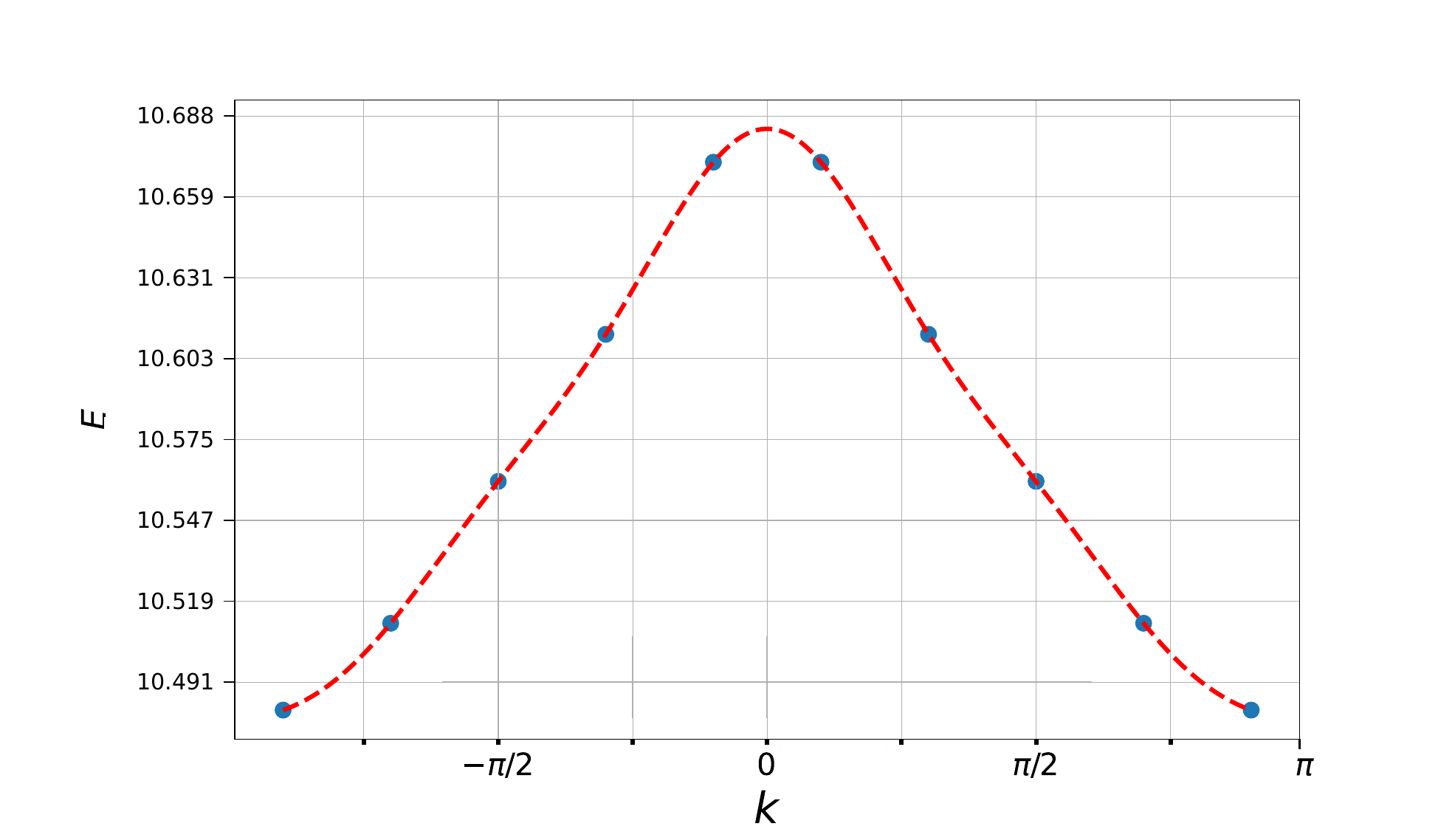} 
\caption{The $\ket{k, -k}$ two-particle scattering state dispersion relation for $N=12$ for $g=1.8$ (\textbf{Left}) and $g=1.5$ (\textbf{Right}). The dispersion should follow the relation given in Eqn. \ref{eqn:dispersion_form}. The red curve provides a fit over those parameters. At $g=1.8$, the dispersion is given by $\alpha + \beta \cos(k)$.  As $g$ is lowered, however, the fit includes higher harmonics.}
\label{fig:bethe_dispersion}
\end{figure}

We first construct the discrete two-particle scattering states for fixed $\sigma_{k} \ll |k_{1}|, |k_{2}|$
\begin{equation}
\begin{aligned}
\psi(k_1, k_2, r_1, r_2) = \frac{1}{\mathcal{N}^{2}} \sum_{k, k'}\left( e^{-(k - k_1)^{2}/2\sigma^{2}_{k}}e^{i r_1 k}\right) \left(e^{-(k' - k_2)^{2}/2\sigma^{2}_{k}} e^{i r_2 k'} \right)\ket{k, k'}
\label{eqn:gaussian_product_state}
\end{aligned}
\end{equation} 
The momentum states $\ket{k, k'}$ are built from two-plaquette excitations. To make a well-defined product state on the plaquette chain, the momentum state should exclude adjacent- and same-plaquette excitations. The ansatz to exclude neighboring plaquette excitations is the Bethe ansatz \cite{Bethe1931}, which takes the form
\begin{equation}
\begin{aligned}
\braket{n_1, n_2 | k_1, k_2} = A\left(e^{-i(k_1 n_1 + k_2 n_2)} + \phi(k_1, k_2)e^{-i(k_1 n_2 + k_2 n_1)}\right)
\end{aligned}
\end{equation} 
where $\phi(k_1, k_2)$ is a phase picked up from the scattering with the exclusion boundary condition. Trivially, the boundary condition without periodicity is given by $\braket{n_1, n_1 +1 | k_1, k_2} = 0$, for which the phase degree of freedom takes the form 
\begin{equation}
\begin{aligned}
\phi(k_1, k_2) = -e^{-i (k_2 - k_1)}
\end{aligned}
\end{equation} 
The periodicity condition for $N$ plaquettes with scattering gives us that 
\begin{equation}
\begin{aligned}
\phi(k_2, k_1)&\braket{n_1 + N, n_2 | k_1, k_2} = \braket{n_1, n_2 | k_1, k_2} \implies \\
&e^{-i k_1 N}\phi(k_1, k_2) = 1 \\ 
&e^{-i k_2 N} \phi(k_2, k_1) = 1  
\end{aligned}
\end{equation} 
Thus, the quantization condition is given as the set of equations 
\begin{equation}
\begin{aligned}
k_1 N - (k_1 - k_2) = (2 m_1 + 1)\pi \\
k_2 N - (k_2 - k_1) = (2 m_2 + 1)\pi
\end{aligned}
\end{equation} 
The independent set of solutions for $k_1$ and $k_2$ is set by the total momentum $k_1 + k_2 = K = \left(2 m_{K} + 2\right)\pi/N$. The quantization condition is $\left(\left(2m + 1\right)\pi - K\right) / \left(N - 2 \right)$. In terms of total and relative momenta, 
\begin{equation}
\begin{aligned}
&K = \frac{2 n \pi}{N}, \\
&\tilde{k} = k_1-k_2 = \frac{2 m \pi}{N-2}, \\
\qquad &n,m\in\mathbb{Z}, \qquad m\neq 0,
\qquad n+m \text{ odd}.
\end{aligned}
\end{equation}
The particles are identical, so $\left(k_1, k_2\right)$ and 
$\left(k_2, k_1\right)$ should be identified. If we take $N$ even, then for $n$ even, there are $(N-2)/2$ unique $m$. For $n$ odd, there are $(N-4)/2$ unique $m$ (since $m = 0$ is excluded). The counting for all states gives $N (N-3)/2$ unique states (this also holds for $N$ odd). This matches with the counting of the two-plaquette separated position basis states. At leading order in perturbation theory, the energies of these states are given by the addition of the dispersions for the momenta $k_1$
and $k_2$, namely $\omega(k_1, k_2) = A + B\left(\cos(k_1) + \cos(k_2) \right)$. In the intermediate-coupling regime, however, the dispersion departs from this. This could be due to a combination of unsuppressed higher-order hopping terms and truncation-induced vacuum mixing.  Fig.~\ref{fig:bethe_dispersion} shows this effect.

\begin{figure}[H]
\centering
\includegraphics[width=0.30\textwidth]{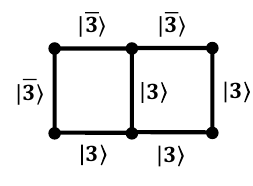}
\caption{A triplet molecule ($\Box_{3, \times 2, 3}$) state at strong coupling. The momentum modes of this particle are made from superpositions of this state and its conjugate.}
\label{fig:triplet}
\end{figure} 

Given the quantization conditions and ansatz, a wave-packet over general $(k_1, k_2)$ would be tedious to construct. For pairs $\left(k_1, k_2\right)$ such that $\phi(k_1, k_2) = 1$, the Bethe ansatz reduces to a decoupled product state. Thus, we will consider tight Gaussians and scattering states with $(k_1, k_2) = (\pi/2, -\pi/2)$.  With these scattering states, we ask what states are produced in a collision of wave-packets. We seek a resonance of the form found in the Friedrichs model, where a discrete bound state couples to an approximate continuum of two-particle scattering states. A natural candidate is the triplet molecule ($\Box_{3, \times 2, 3}$, Fig. \ref{fig:triplet}), as its electric Casimir is close to the two $\Box_{3}$ electric Casimirs. We find that for $g \in [1.2, 1.4]$, the scattering states have avoided crossings with $\Box_{3, \times 2, 3}$.

\subsubsection{Characterization of the Resonance with the $\Box_{3, \times 2, 3}$}
For the $K=0$ sector, the restricted Hilbert space contains a single bound state that interacts with the scattering states. The continuum Hamiltonian for the Friedrichs model takes the form

\begin{equation}
\begin{aligned}
H = \omega_{0} \ket{0}\bra{0} + \int d{k} \, \omega_{k} \ket{k} \bra{k} + \int dk \, [g^{*}(k) \ket{k}\bra{0} + g(k) \ket{0}\bra{k}]
\end{aligned}
\end{equation} 
\noindent where $\ket{0}$ is the bound state, $\ket{k}$ are the scattering states, and $g(k)$ is the form factor for the interactions between the two. This interaction term is the potential $V$. For the same theory, but now discretized into $N$ sites, the form factor changes accordingly. The scattering states are labeled by $(N-2)/2$ $k$, defined as 
\begin{equation}
    k_1 = -k_2 \equiv k,
    \qquad
    k = \frac{(2j+1)\pi}{N-2},
    \qquad
    j = 0,1,\ldots
\end{equation}
We then define the matrix element between the bound state and scattering state with 
\begin{equation}
\begin{aligned}
\mathcal{V}_{k} \equiv \braket{0 | V | k}_{\text{box}
} = \sqrt{\frac{2 \pi}{N-2}} g(k)
\end{aligned}
\end{equation}
The factor $\sqrt{2 \pi/(N-2)}$ comes from matching to the continuum. Namely, 
\begin{equation}
\begin{aligned}
&\sum_{k} \to \left(\frac{2\pi}{N-2}\right)^{-1} \int d k \\
V_{d} = \left(\frac{2 \pi}{N-2}\right)^{-1} &\int d k \, \, \left(g^{*}_{d}(k) \ket{k}_{d}\bra{0} + g_{d}(k) \ket{0}\bra{k}_{d}\right) \\
\ket{k}_{d} = &\left( \sqrt{\frac{2 \pi}{N-2}} \right)\ket{k}_{c} \implies g(k)_{d} = \left( \sqrt{\frac{2 \pi}{N-2}} \right)g(k)_{c} 
\end{aligned}
\end{equation}
\begin{figure}[t]
\centering
\includegraphics[width=0.80\textwidth]{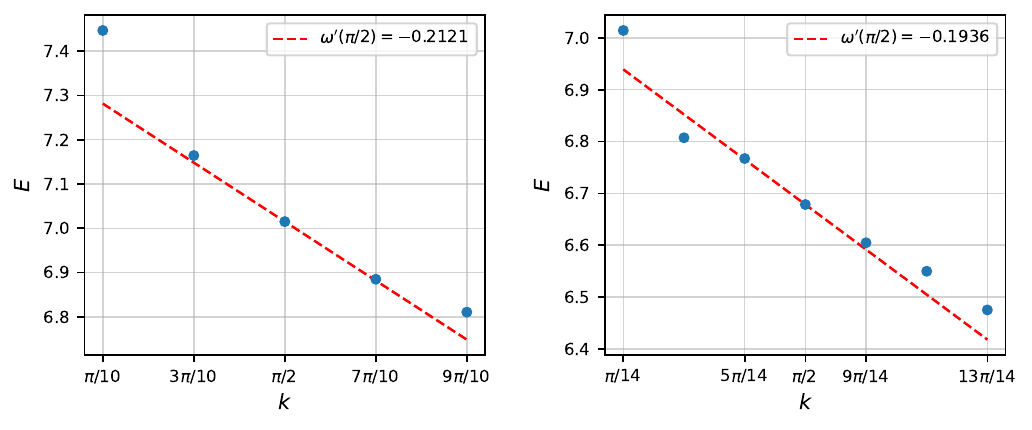} 
\caption{The slope of the dispersion relation $\omega(k, -k)$ at $k = \pi/2$; \textbf{Left:} $N=12$, \textbf{Right:} $N=16$.}
\label{fig:k=0_slope}
\end{figure}
\begin{figure}[t]
\centering
\hfill 
\hskip -1.5em
\includegraphics[width=0.50\textwidth]{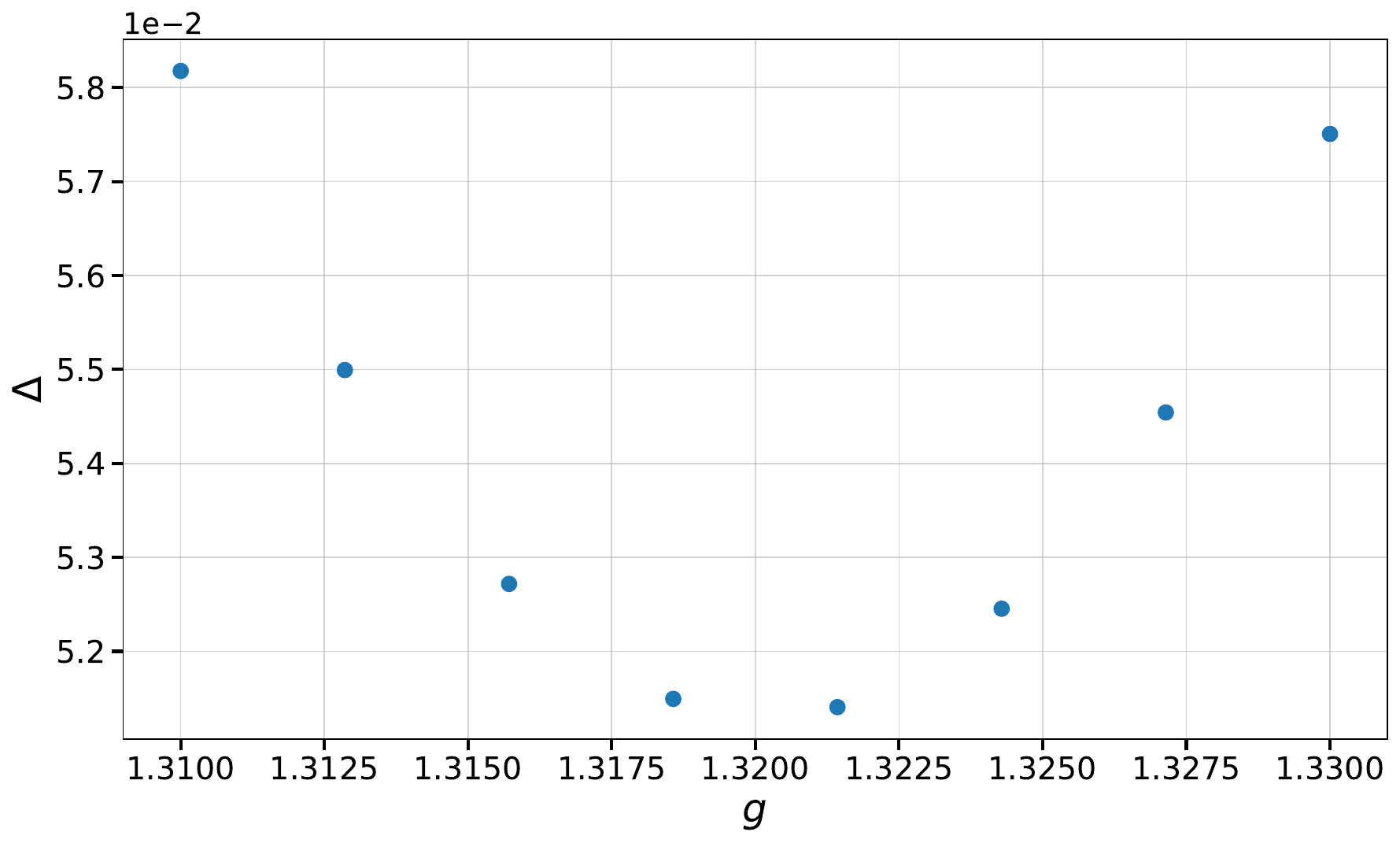} %
\includegraphics[width=0.50\textwidth]{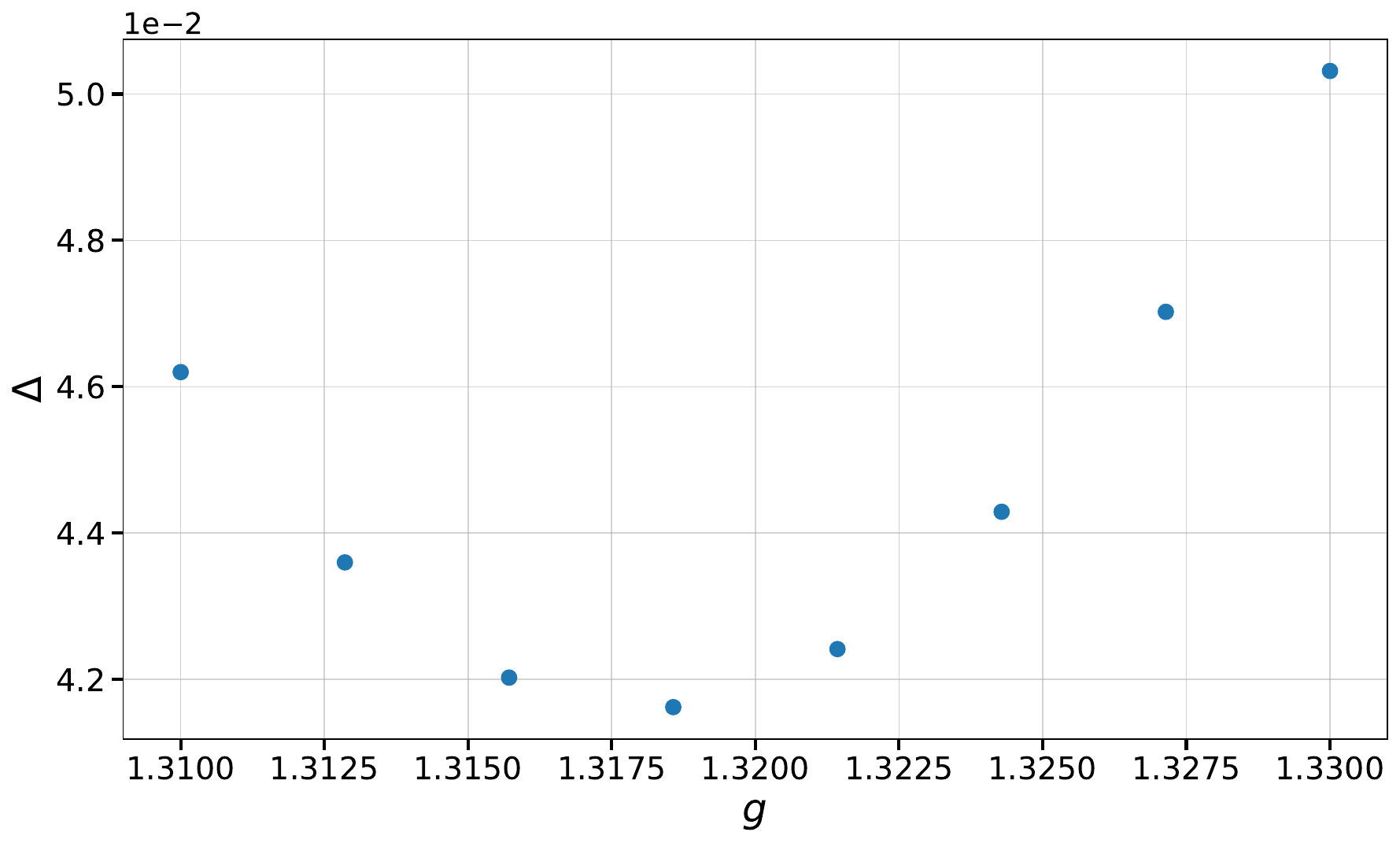}
\caption{$\Delta_{\text{AC}}$ for $N=12$ and $N=16$.}
\label{fig:Delta_AC}
\vspace{-0.0cm}
\end{figure}
\begin{figure}[t]
\centering
\hfill 
\hskip -1.5em
\includegraphics[width=0.50\textwidth]{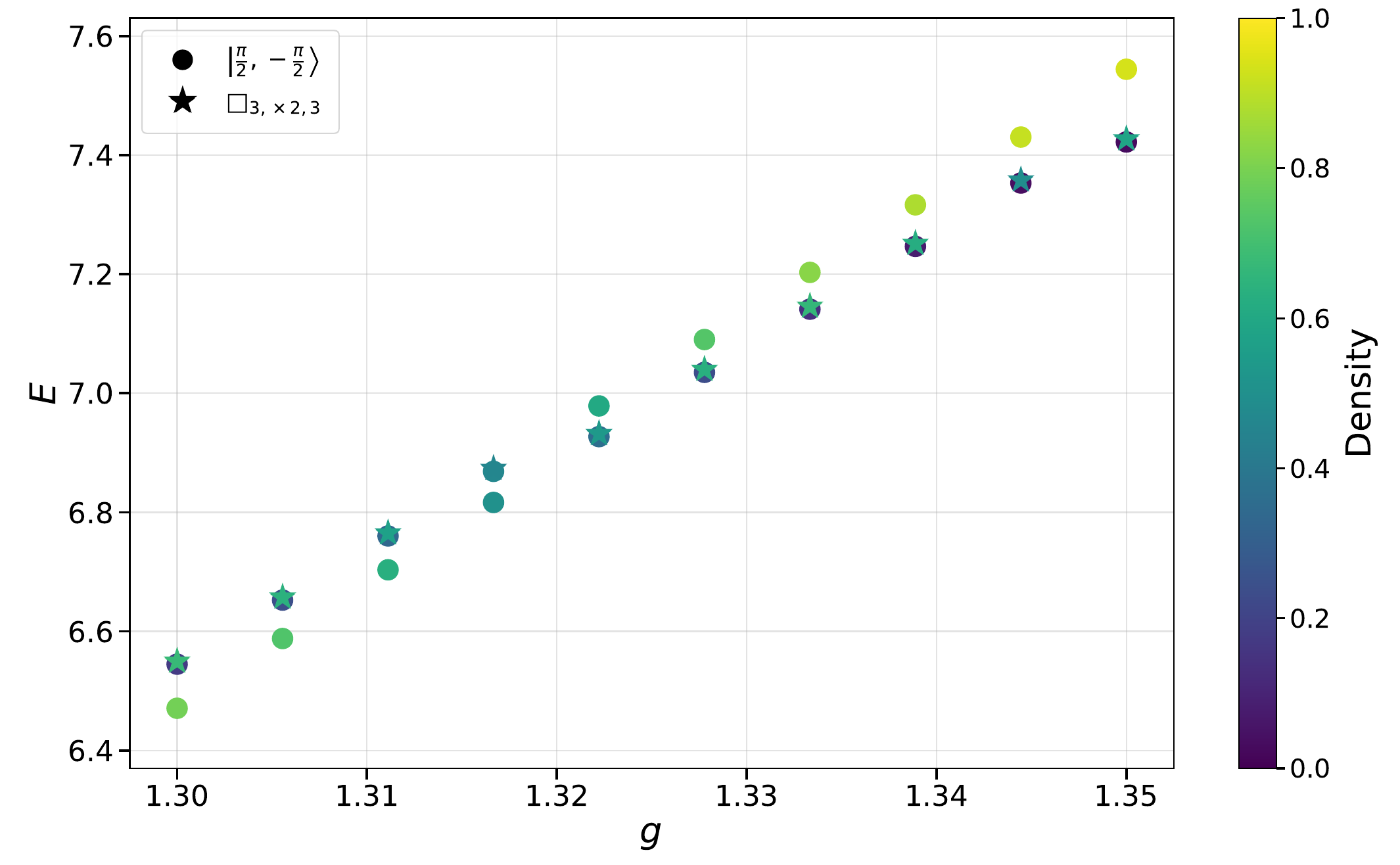} %
\includegraphics[width=0.50\textwidth]{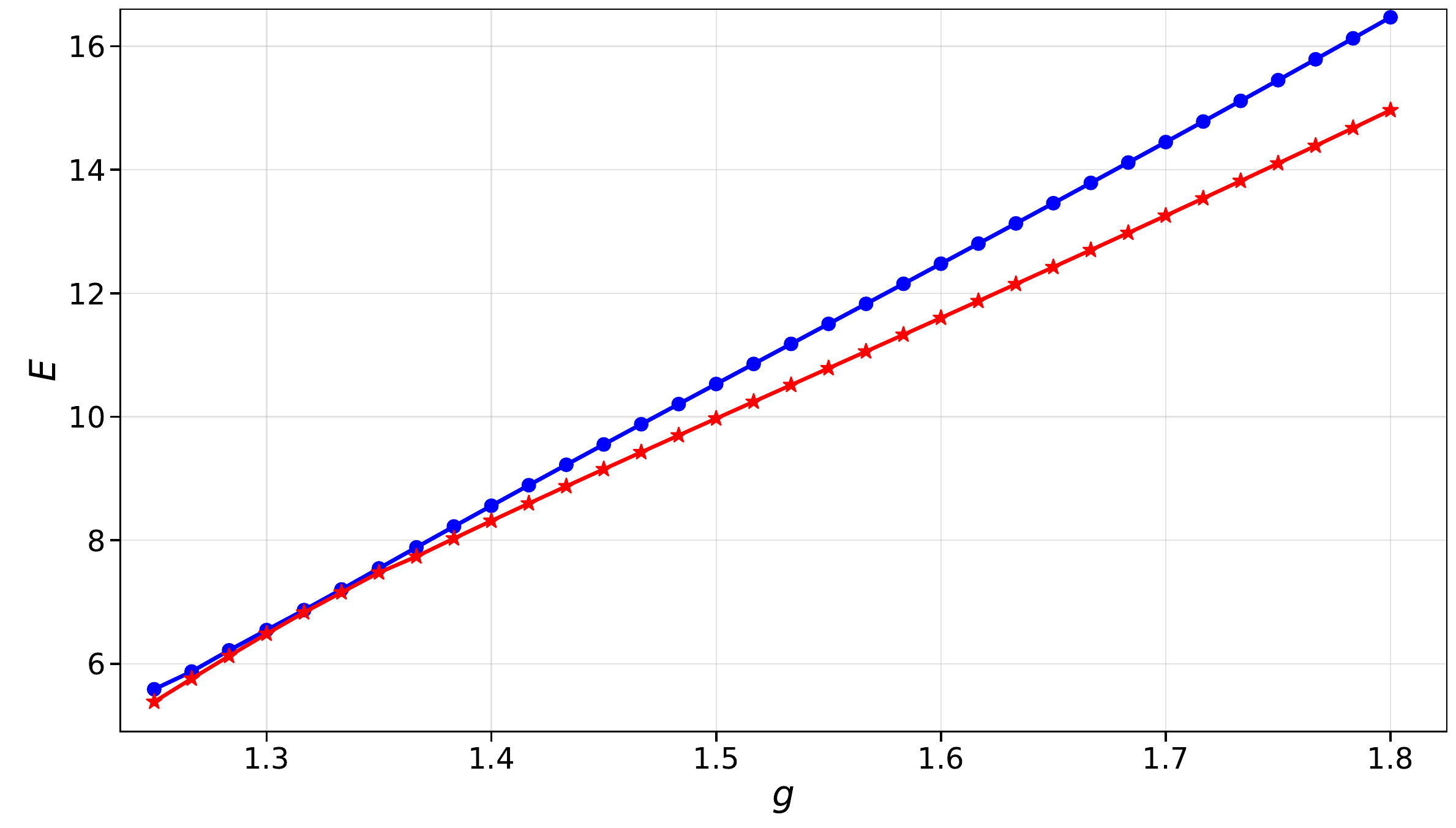} %
\caption{$\ket{\frac{\pi}{2}, -\frac{\pi}{2}}$ scattering state and $\Box_{3, \times 2, 3}$ eigenstate support as a function of $g$ for $N=12$ (\textbf{Left}). Eigenstates in the crossing (\textbf{Right}).}
\label{fig:wave-packet_support}
\end{figure} 
where the subscript $d$ refers to the discrete coupling and subscript $c$ refers to the continuous coupling. When the discrete state energy lies close to a scattering state energy (the resonant state), the Hamiltonian can be approximated as a $2 \, \times \, 2$ model with the other modes integrated out. The Hamiltonian at a particular energy is 
\begin{center}
$ H_{*}\!\left(E \right) = 
\begin{pmatrix}
\omega_{0} + \delta \omega_{0}(E) & \mathcal{V}_{n_{*}} \\
\mathcal{V}^{*}_{n_{*}} & \omega_{n_{*}} \\
\end{pmatrix},$
\end{center}
with 
\begin{equation}
\begin{aligned}
\delta \omega_{0}\!\left( E\right) = \sum_{n \neq n_{*}} \frac{|\mathcal{V}_{n}|^{2}}{E - \omega_{n}}
\end{aligned}
\end{equation}
coming from the self-energy correction due to interaction with all scattering energies. At the resonant energy $E = E_R$, we expect that $\omega_{0} + \delta \omega_{0}(E_R) =  \omega_{n_{*}} = E_{R}$. We may then solve for the eigenvalues of $H_{*}(E)$ in terms of $E_{R}$ with the characteristic equation
\begin{equation}
\begin{aligned}
(E - E_R)\left[E - E_{R} - \left(\delta \omega_{0}(E) - \delta\omega_{0}(E_R)\right) \right] - |V_{n_{*}}|^{2} = 0
\end{aligned}
\end{equation}
where we define the two solutions near the selected avoided crossing as $E_{\pm}$ and define an avoided crossing gap $\Delta_{\text{AC}}$ as $|E_{+} - E_{-}| \equiv \Delta_{\text{AC}}$. 
To leading order,
\begin{equation}
\delta\omega_0(E)-\delta\omega_0(E_R)
\simeq
(E-E_R)\delta\omega_0'(E_R),
\label{eq:self_energy_variation}
\end{equation}
where
\begin{equation}
\delta\omega_0'(E_R)
=
-\sum_{n\neq n_*}
\frac{|\mathcal{V}_n|^2}{(E_R-\omega_n)^2}.
\label{eq:self_energy_derivative}
\end{equation}
If this sum is dominated by a few nearby off-resonant modes
with energy separation $\delta\omega$ and couplings comparable
to the resonant coupling,
$|\mathcal{V}_n|\sim|\mathcal{V}_{n_*}|\sim|\mathcal{V}|$,
then
\begin{equation}
    |\delta\omega_0'(E_R)|
    \sim
    \frac{|\mathcal{V}|^2}{(\delta\omega)^2}.
\end{equation}
Within the avoided-crossing window,
$|E-E_R|\sim|\mathcal{V}_{n_*}|$,
the omitted shift therefore has magnitude
\begin{equation}
\left|
\delta\omega_0(E)-\delta\omega_0(E_R)
\right|
\sim
|\mathcal{V}|
\frac{|\mathcal{V}|^2}{(\delta\omega)^2}
=
\frac{|\mathcal{V}|^3}{(\delta\omega)^2}.
\label{eq:omitted_self_energy_shift}
\end{equation}
For the energy-independent $2\times2$ approximation to resolve
the avoided-crossing splitting accurately, this correction
must be small compared with
$\Delta_{\mathrm{AC}}\sim|\mathcal{V}|$:
\begin{equation}
\frac{|\mathcal{V}|^3}{(\delta\omega)^2}
\ll |\mathcal{V}|
\quad\Longleftrightarrow\quad
|\mathcal{V}|^2\ll(\delta\omega)^2
\quad\Longleftrightarrow\quad
|\mathcal{V}|\ll\delta\omega.
\label{eq:isolated_crossing_condition}
\end{equation}
With this condition, we may ignore the correction in Eq. \ref{eq:self_energy_variation} and get that 
\begin{equation}
\begin{aligned}
&\ket{E_{\pm}} \approx \frac{1}{\sqrt{2}} \left(\ket{0} \pm e^{-i \, \text{arg} \left(V_{n_{*}}\right)} \ket{k_{n_{*}}} \right) \\
&E_{\pm} \approx \omega_{n_{*}} \pm |\mathcal{V}_{n_{*}}|
\end{aligned}
\end{equation}
with the avoided crossing gap given as $\Delta_{\text{AC}} \simeq 2 |\mathcal{V}_{n_{*}}|$. With this definition, the resonance width as given by Fermi's golden rule can be written in terms of the avoided crossing gap as 
\begin{equation}
\begin{aligned}
\Gamma \simeq 2 \pi |g(k_{n_{*}})|^{2} \rho(\omega_{n_{*}}) = \frac{(N-2) \Delta^{2}_{\text{AC}} \rho_{K=0}(\omega_{n_{*}})
}{4} \quad \text{with} \\
\rho_{K=0}(\omega) = \sum_{k^{*}} \, \frac{1}{|d \omega(k, -k) / dk|}
\label{eqn:gamma}
\end{aligned}
\end{equation}
\begin{table}[t]
    \centering
    \caption{Extracted resonance quantities for the $N=12$ and $N=16$ plaquette chains. 
    Here $g_R$ is the coupling at resonance, $k_*$ is the resonant scattering momentum, 
    $E_R$ is the resonance energy, $\Delta_{\mathrm{AC}}$ is the avoided-crossing gap, 
    and $\rho(\omega_{n_{*}})$ is the continuum density of states evaluated at the resonance energy.}
    \label{tab:resonance_parameters}
    \begin{tabular}{c c c c c c c c}
        \hline
        $N$ 
        & $g_R$ 
        & $k_*$ 
        & $E_R$ 
        & $\Delta_{\mathrm{AC}}$ 
        & $\left|\dfrac{d\omega}{dk}\right|_{k_*}$ 
        & $\rho(\omega_{n_{*}})$ 
        & $\Gamma$ \\
        \hline
        12 & 1.32 & $\pi/2$ & 7.0 & 0.051 & 0.2121 & 4.71 & 0.031 \\
        16 & 1.318 & $\pi/2$ & 6.7 & 0.042 & 0.1936 & 5.16 & 0.032 \\
        \hline
    \end{tabular}
\end{table}

We estimate the continuum resonant width for the $\Box_{3, \times 2, 3}$ resonance for the $N=12$ and $N=16$ plaquette chains through measurement of $\Delta_{\text{AC}}$ and $\rho(\omega)$ at the resonant coupling. With $|d\omega(k, -k)/dk|$ in Fig.~\ref{fig:k=0_slope} and $\Delta_{AC}$ in Fig.~\ref{fig:Delta_AC}, $|\mathcal{V}|$ is smaller than the scattering state energy splittings. Eq. \ref{eqn:gamma} therefore provides a leading-order approximation to $\Gamma$. Table \ref{tab:resonance_parameters} shows the full resonance quantities between $N=12$ and $N=16$. The biggest discrepancy is the measured $|d\omega(k, -k)/dk|$, which is expected with discretization of the momenta and the presence of other resonances. The more useful quantity for comparing different chain lengths $N$ is the ratio of the respective $\Delta_{AC}$, which should match the ratio of the respective $1/\sqrt{N-2}$. We find that 
\begin{equation}
\begin{aligned}
\frac{\left(\Delta_{\text{AC}} \times \sqrt{N-2}\right)_{N=12}}  {\left(\Delta_{\text{AC}} \times \sqrt{N-2}\right)_{N=16}} \approx 1.03.
\end{aligned}
\end{equation}
\begin{figure}[t]
\centering
\hfill
\includegraphics[width=0.50\textwidth]{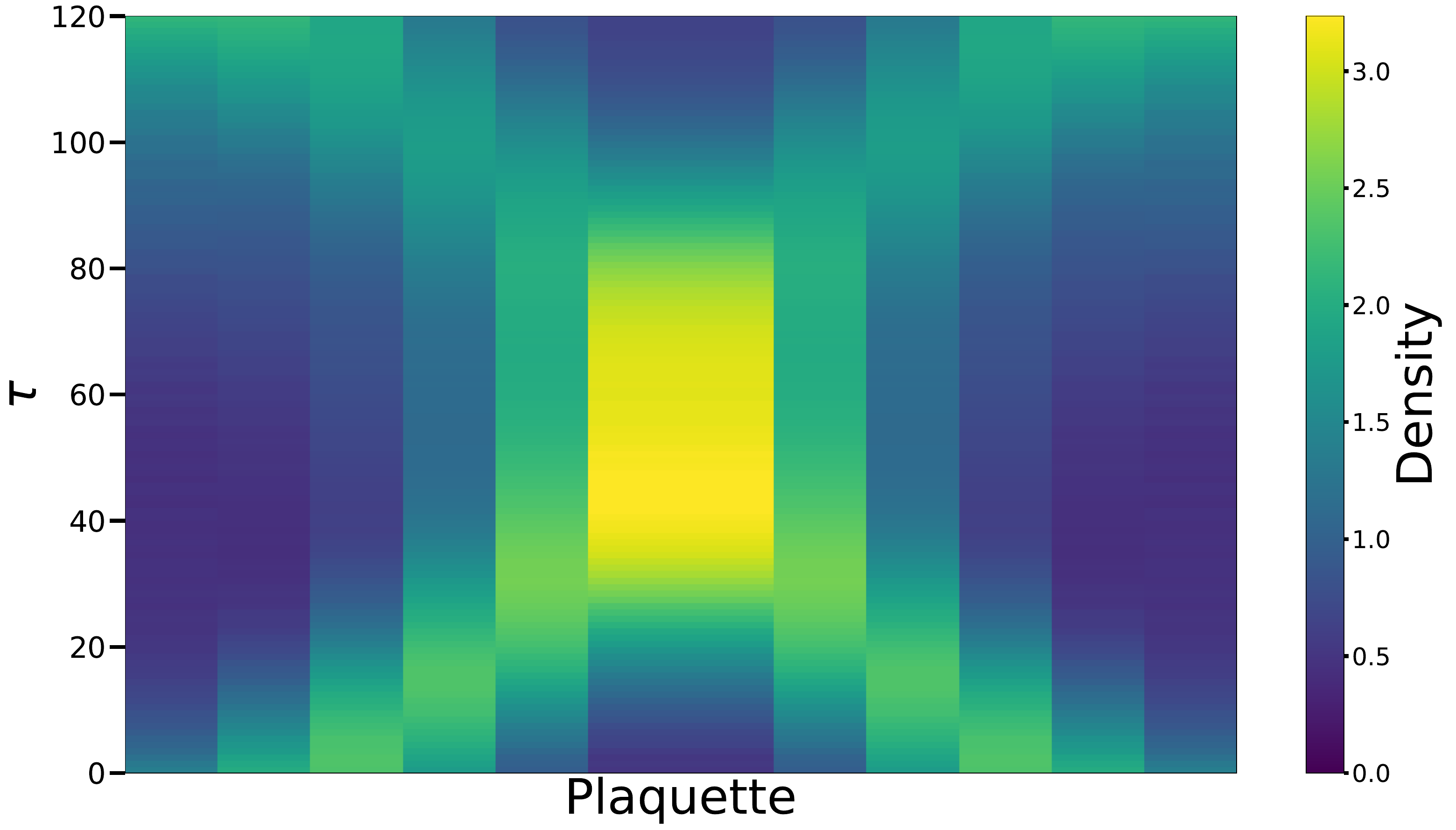} %
\hskip -0.5em
\includegraphics[width=0.50\textwidth]{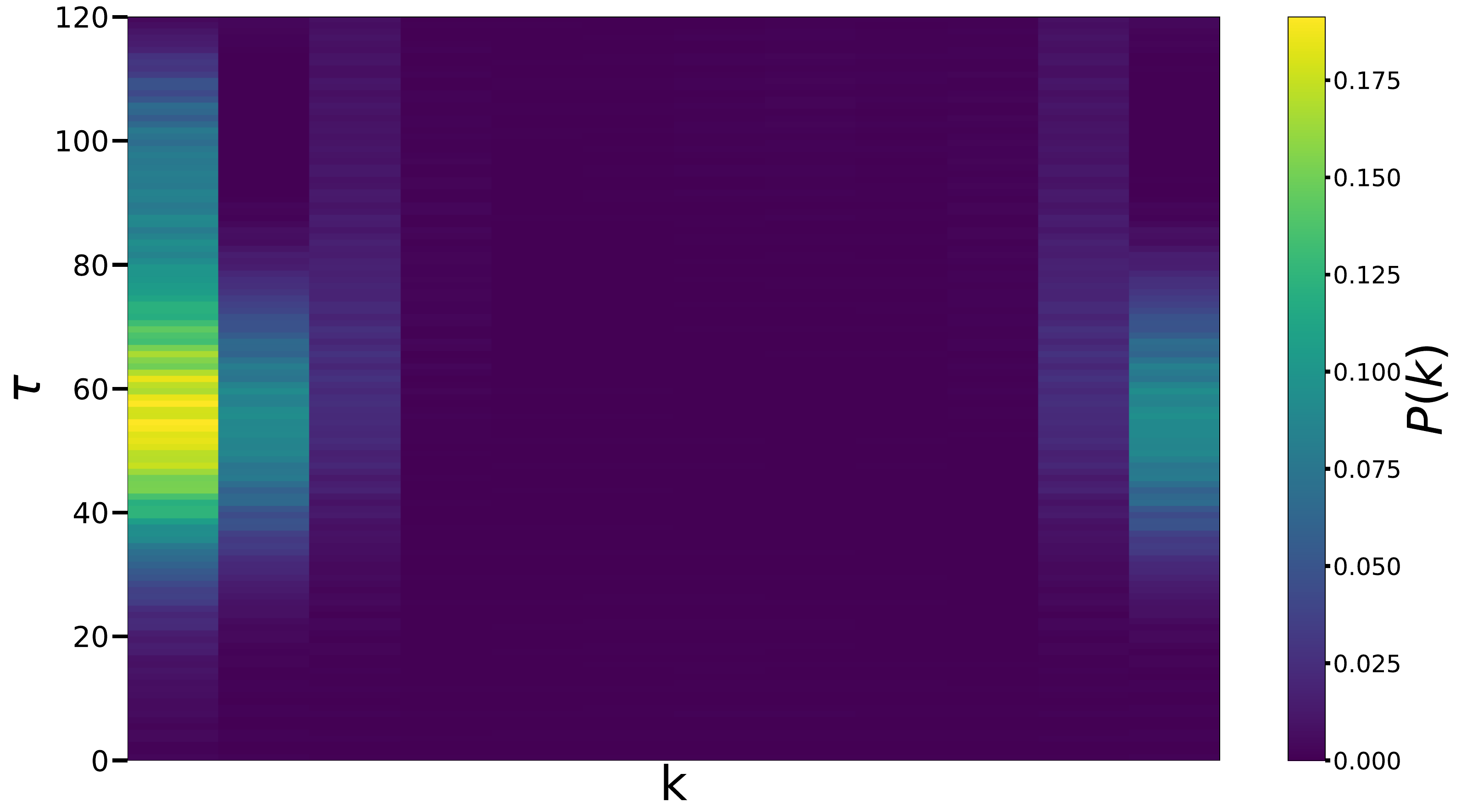}
\caption{\textbf{Left:} $\left(k_1, k_2 \right) = \left(\pi/2, -\pi/2 \right)$ collision with $ N = 12, \, N \sigma_{k} / 2
\pi = 1.0$, \textbf{Right:} Momentum space production of the raw molecule as a function of time.}
\label{fig:collisions}
\end{figure} 
\subsubsection{Defining Time-Dependent Resonance Observables}
The spectral analysis indicates that the resonance at $k=\pi/2$ is well approximated by the $2 \times 2$ Friedrichs model and we can get a precise measurement of $\Delta_{\text{AC}}$. However, such analysis does not scale to weak coupling, more complex glueball collisions, or $N \gg 1$; more general characterization requires extracting resonant signatures from real-time simulations. Extracting the same numerical quantities requires defining the right observables $\braket{O}$. One interesting time-domain observable is the Wigner time-delay $\tau_{W}$ \cite{Wigner1955,Gustafson2021TimeDelay}. A clear time delay is seen by tracking the energy density of each wave-packet before and after the collision (Fig.~\ref{fig:collisions}). We treat this as a qualitative signature. A related observable from which we can more easily extract accurate microscopic quantities is the Rabi oscillation of the scattering state responsible for the resonance.
If we diabatically\footnote{A ``diabatic" ramp here means a large Landau-Zener transition probability at the avoided crossing, but a small probability at other crossings.} ramp the scattering state down to the resonant coupling, then, to a good approximation, with $\alpha, \beta 
\sim 1/\sqrt{2}$, the state is
\begin{figure}[H]
\begin{center}
\hspace{-1cm}
\hfill
\includegraphics[width=0.50\textwidth]{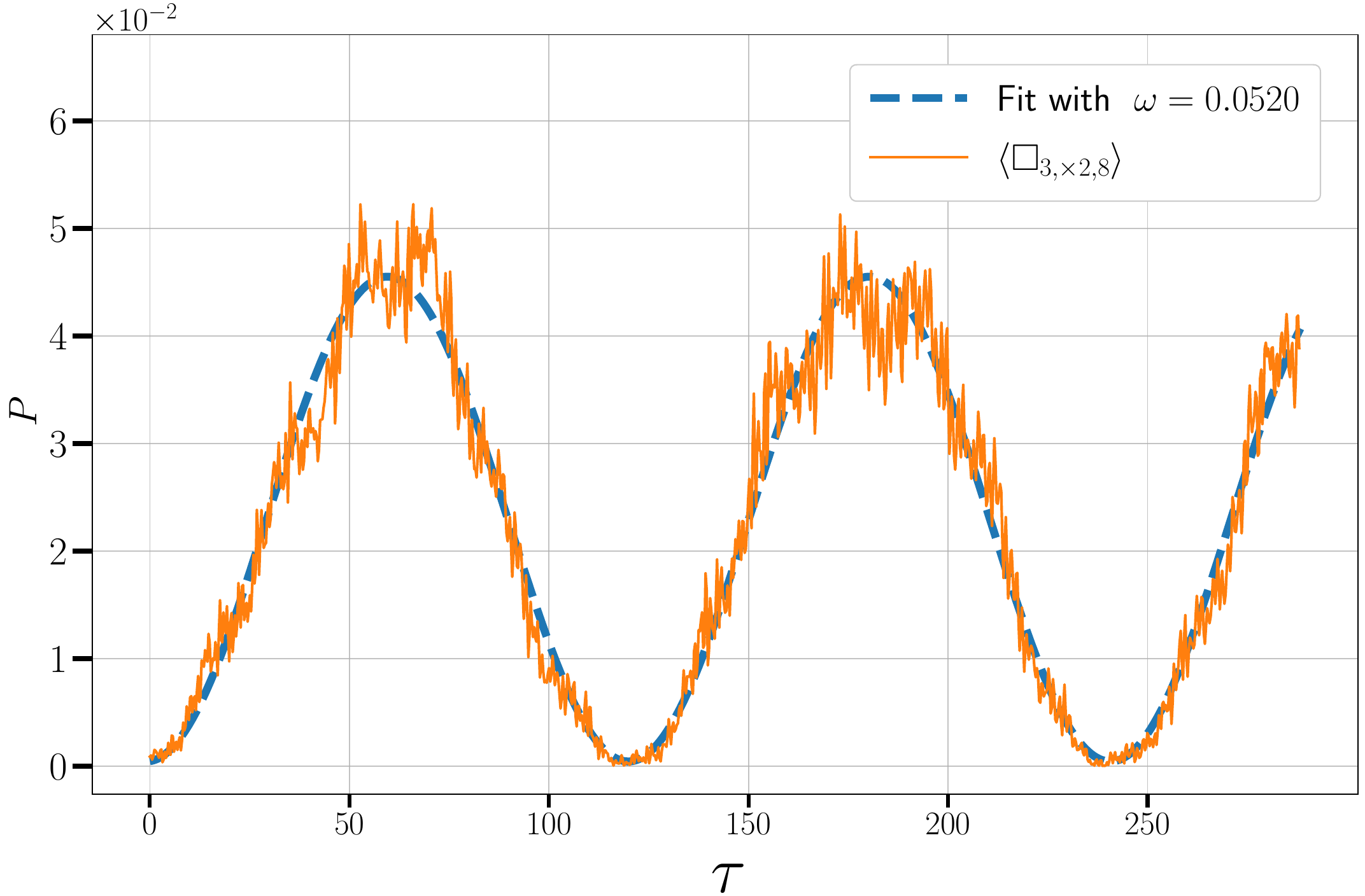} 
\includegraphics[width=0.50\textwidth]{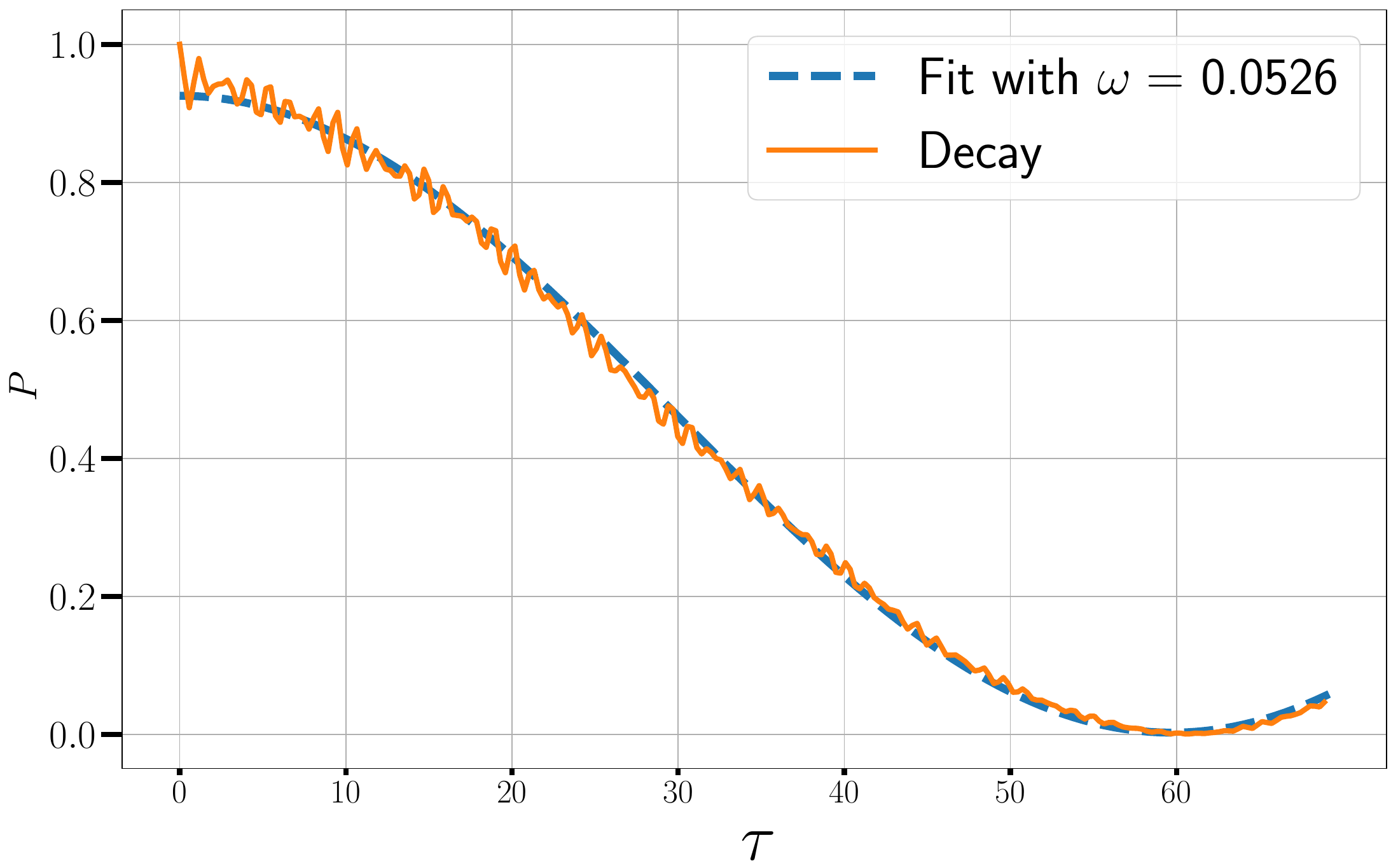} %
\vspace{-0.5cm}
\end{center}
\caption{Measurement of $\omega = \Delta_{\text{AC}}$ for $N=12$: resonance of $\Box_{3, \times 2, 3}$ through a state projection (\textbf{Left}) and survival probability (\textbf{Right}).}
\vspace{-0.25cm}
\label{fig:rabi}
\end{figure}
\begin{equation}
\begin{aligned}
\ket{k_1, k_2} = \alpha \ket{E_+} + \beta \ket{E_-} + \delta \ket{(k_1, k_2)_{\text{small}}}
\end{aligned}
\end{equation}
The survival probability is given by $|\braket{k_1, k_2|e^{-i H t} | k_1, k_2}|^{2}$
\begin{equation}
\begin{aligned}
|\braket{k_1, k_2|e^{-i H t} | k_1, k_2}|^{2} = &\left|\left(|\alpha|^{2}e^{iE_{+} t} + |\beta|^{2}e^{iE_{-} t} + O\left(\delta^{2} \right) \right)\right|^{2} \\
= &\left(|\alpha|^{4} + |\beta|^{4} + 2|\alpha|^{2} |\beta|^{2}\cos\left(\Delta_{\text{AC}} \, t \right)\right) + O\left(\delta^{2} \right)
\label{eqn:rabi_equation}
\end{aligned}
\end{equation}
The period of the oscillation gives $\Delta_{\text{AC}}$. However, preparing a measurement of the survival probability requires not only a diabatic state preparation for the initial scattering state, but also for the final scattering state for readout. We may avoid the second ramp by projecting the evolved scattering state onto an unramped state in the $K=0$ sector. The simplest choice is another high-energy state far from the bound state:
\begin{equation}
\begin{aligned}
\braket{\psi | e^{-i H t} | k_1, k_2} \approx \frac{1}{\sqrt{2}} \left(e^{-iE_+ t }\braket{\psi | E_{+}} + e^{-iE_- t }\braket{\psi | E_{-}} \right)  \\
\alpha_{\pm} = \braket{\psi|E_{\pm}}, \quad |\!\braket{\psi | e^{-i H t} | k_1, k_2}\!|^{2} \approx \frac{1}{2}\left(|\alpha_{+}|^{2} + |\alpha_{-}|^{2} + 2|\alpha_{+}||\alpha_{-}| \cos\left(\Delta_{\text{AC}}t + \phi \right)\right)
\end{aligned}
\end{equation}
Both methods can yield very precise measurements of $\Delta_{\text{AC}}$, as seen in the classical extraction of the $\Box_{3, \times 2, 3}$ gap in Fig.~\ref{fig:rabi}. To achieve this in a quantum simulation, the oscillation would have to be resolved over a time series of simulations, with a large enough number of samples to extract a signal, small enough d$\tau$ to distinguish the curve, and long enough simulation time $\tau$ for high resolution in a frequency decomposition. \\

Thus the steps we will follow to extract  resonance properties from real-time simulations are:
\begin{enumerate}[leftmargin=2cm]
        \item Construct a wave-packet collision and observe a qualitative sign of resonance (time-delay)
        \item Measure the $\Delta_{\text{AC}}$ by looking at the Rabi oscillation of the scattering state \\
\end{enumerate} 

We illustrate both procedures through toy model simulations and demonstrate the wave-packet collision on quantum hardware.

\section{Observing $SU(3)$ Resonance on a Quantum Device}

\subsection{Toy Model}
We consider a toy model of $SU(3)$ glueball propagation and interactions using the KS Hamiltonian in the electric basis. The goal is to construct a minimal model representative of the full theory that can run on near-term devices. To this end we consider the production of a more massive glueball state from the collision of the lightest glueball states on a chain of plaquettes with PBCs. We consider a Hilbert space where the only plaquette states are low-energy glueball basis states $+$, this high-energy glueball state $e$, and the vacuum. The effective Hamiltonian is defined on adjacent sites and includes an interaction term for the $+, e$ states and a hopping term for the $+$ states:
\begin{equation}
\begin{aligned}
H_{\mathrm{eff}} = \sum_{i} \Big[&
\alpha_{1}
\left(
\ket{+_{i}0_{i+1}}\bra{0_{i}+_{i+1}}
+
\ket{0_{i}+_{i+1}}\bra{+_{i}0_{i+1}}
\right)
\\
&+
\alpha_{2}
\left(
\ket{e_{i}0_{i+1}}\bra{+_{i}+_{i+1}}
+
\ket{+_{i}+_{i+1}}\bra{e_{i}0_{i+1}}
\right)
\\
&+
\epsilon_{+}
\ket{+_{i}}\bra{+_{i}}+
\epsilon_{e}
\left(
\ket{e_{i}}\bra{e_{i}}
\right)
\Big].
\end{aligned}
\end{equation}

\begin{table}[t]
\centering
\vspace*{-20pt}

\begin{tabular}{|l|l|l|}
\hline
\multicolumn{3}{|c|}{\textbf{Local Plaquette States}} \\ \hline
\textbf{State} & \textbf{Description} & \textbf{Qubit Encoding} \\ \hline
$\ket{\mathbf{0}}$ 
& Vacuum 
& $\ket{00}$ \\ \hline

$\ket{\mathbf{+}}$ 
& Low-energy glueball 
& $\ket{10}$ \\ \hline

$\ket{\mathbf{e}}$ 
& High-energy glueball 
& $\ket{01}$ \\ \hline

\multicolumn{3}{|c|}{\textbf{Interactions}} \\ \hline
\textbf{Initial State} & \textbf{Final State} & \textbf{Qubit Encoding} \\ \hline

$\ket{\mathbf{+}}_{n}\ket{\mathbf{+}}_{n+1}$
&
$\ket{\mathbf{e}}_{n}\ket{\mathbf{0}}_{n+1}$
&
$\ket{1010}\leftrightarrow\ket{0100}$
\\ \hline

$\ket{\mathbf{+}}_{n}\ket{\mathbf{+}}_{n+1}$
&
$\ket{\mathbf{0}}_{n}\ket{\mathbf{e}}_{n+1}$
&
--- 
\\ \hline

\multicolumn{3}{|c|}{\textbf{Hopping and Electric Energy Terms}} \\ \hline
\textbf{Initial State} & \textbf{Final State} & \textbf{Qubit Encoding} \\ \hline

$\ket{\mathbf{+}}_{n}\ket{\mathbf{0}}_{n+1}$
&
$\ket{\mathbf{0}}_{n}\ket{\mathbf{+}}_{n+1}$
&
$\ket{1000}\leftrightarrow\ket{0010}$
\\ \hline

$\ket{\mathbf{+}}_{n}$
&
$\ket{\mathbf{+}}_{n}$
&
$\ket{10}\leftrightarrow\ket{10}$
\\ \hline

$\ket{\mathbf{e}}_{n}$
&
$\ket{\mathbf{e}}_{n}$
&
$\ket{01}\leftrightarrow\ket{01}$
\\ \hline

\end{tabular}

\caption{
Complete set of local states and transitions for the toy Hamiltonian,
together with their qubit encodings. The plaquette states are the vacuum
$\ket{0}$, low-energy glueball state $\ket{+}$, and high-energy glueball
state $\ket{e}$. Only one of the two
$\ket{++}\leftrightarrow\ket{e0},\ket{0e}$ interaction channels is
explicitly encoded in the circuit in order to reduce the gate cost,
without changing the relevant dynamics.
}
\label{tab:local_encoding}

\end{table}

For a quantum simulation, the most scalable encoding is a local one. We therefore choose a plaquette-local encoding where each plaquette is represented by two qubits (for 3 states). Table~\ref{tab:local_encoding} shows the encoding. It should be noted that this construction and encoding excludes states that form from adjacent plaquette excitations, which do appear in the actual theory. This toy model is still a good representative model for the dynamics that do occur in the full theory. 

\subsection{Circuit Decomposition}
\subsubsection{Hamiltonian Evolution}
\label{evolution_circuits}
\begin{figure}[H]

\begin{center}
$e^{-i \theta \left(XX + YY\right)}\equiv$ 
\begin{quantikz}
\lstick{$q_0$} & \gate{H} & \gate[2]{R_{ZZ}\left(2\theta \right)} & \gate{H} & \gate{HS^{\dagger}} & \gate[2]{R_{ZZ}\left(2\theta \right)} & \gate{SH} \\
\lstick{$q_1$} & \gate{H} & \qw & \gate{H} & \gate{HS^{\dagger}} & & \gate{SH}
\end{quantikz}
\caption{Circuit decomposition of the hopping term in terms of two $R_{ZZ}$ rotations.}
\label{fig:hopping_RZZ}
\end{center}
\end{figure}
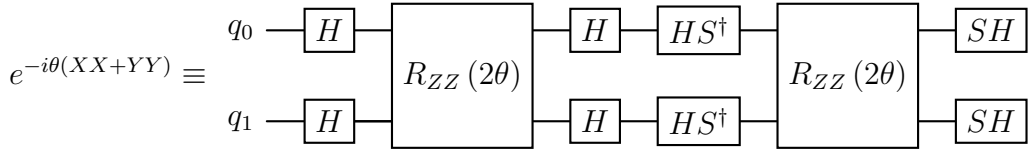

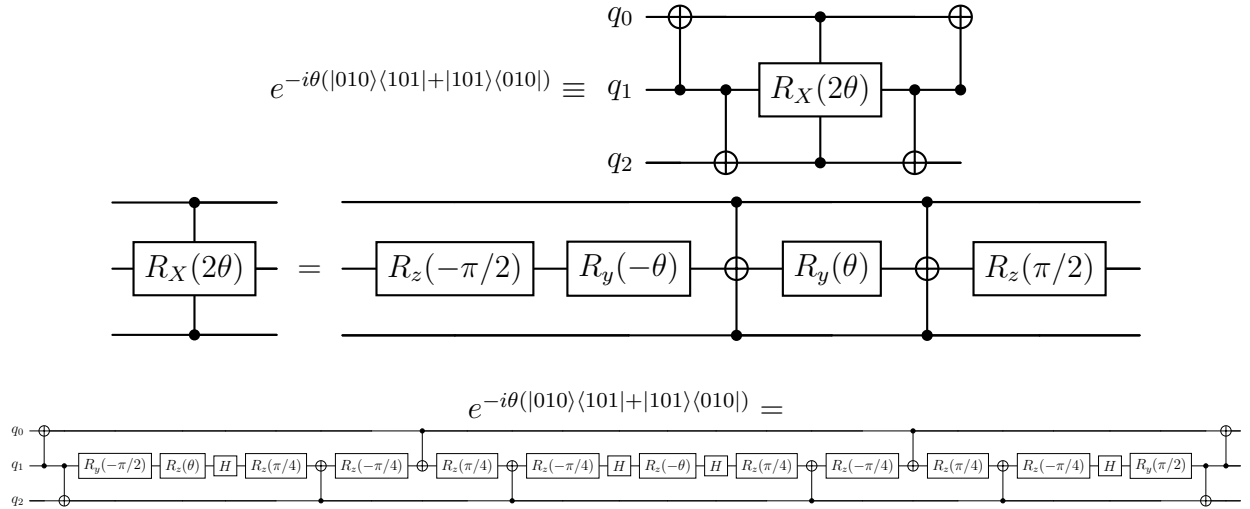
\begin{figure}[H]
\begin{center}
$e^{-i \theta \left(\ket{010}\bra{101} + \ket{101}\bra{010}  \right)}\equiv$ 
\begin{quantikz}[row sep=0.45cm, column sep=0.28cm]
\lstick{$q_0$}
    & \targ{}
    & \qw
    & \ctrl{1}
    & \qw
    & \targ{}
\\
\lstick{$q_1$}
    & \ctrl{-1}
    & \ctrl{1}
    & \gate{R_{X}(2\theta)}
    & \ctrl{1}
    & \ctrl{-1}
\\
\lstick{$q_2$}
    & \qw
    & \targ{}
    & \ctrl{-1}
    & \targ{}
    & \qw
\end{quantikz} \\
\begin{quantikz}[row sep=0.45cm, column sep=0.28cm]
    & \ctrl{1}
    & \qw
\\
    & \gate{R_{X}(2\theta)}
    & \qw
\\
    & \ctrl{-1}
    & \qw
\end{quantikz} = \begin{quantikz}[row sep=0.45cm, column sep=0.45cm]
    & \qw
    & \qw
    & \ctrl{1}
    & \qw
    & \ctrl{1}
    & \qw
    & \qw
\\
    & \gate{R_z(-\pi/2)}
    & \gate{R_y(-\theta)}
    & \targ{}
    & \gate{R_y(\theta)}
    & \targ{}
    & \gate{R_z(\pi/2)}
    & \qw
\\
    & \qw
    & \qw
    & \ctrl{-1}
    & \qw
    & \ctrl{-1}
    & \qw
    & \qw
\end{quantikz} \\
\vspace{0.5cm}
$e^{-i \theta \left(\ket{010}\bra{101} + \ket{101}\bra{010}  \right)}=$ \\
\resizebox{\textwidth}{!}{
\begin{quantikz}[
    row sep=0.48cm,
    column sep=0.25cm
]
%
%
\lstick{$q_0$}
& \targ{}
& \qw
& \qw
& \qw
& \qw
& \qw
& \qw
& \qw
& \ctrl{1}
& \qw
& \qw
& \qw
& \qw
& \qw
& \qw
& \qw
& \qw
& \qw
& \ctrl{1}
& \qw
& \qw
& \qw
& \qw
& \qw
& \qw
& \targ{}
& \qw
\\
%
%
\lstick{$q_1$}
& \ctrl{-1}
& \ctrl{1}
& \gate{R_y(-\pi/2)}
& \gate{R_z(\theta)}
& \gate{H}
& \gate{R_z(\pi/4)}
& \targ{}
& \gate{R_z(-\pi/4)}
& \targ{}
& \gate{R_z(\pi/4)}
& \targ{}
& \gate{R_z(-\pi/4)}
& \gate{H}
& \gate{R_z(-\theta)}
& \gate{H}
& \gate{R_z(\pi/4)}
& \targ{}
& \gate{R_z(-\pi/4)}
& \targ{}
& \gate{R_z(\pi/4)}
& \targ{}
& \gate{R_z(-\pi/4)}
& \gate{H}
& \gate{R_y(\pi/2)}
& \ctrl{1}
& \ctrl{-1}
& \qw
\\
%
%
\lstick{$q_2$}
& \qw
& \targ{}
& \qw
& \qw
& \qw
& \qw
& \ctrl{-1}
& \qw
& \qw
& \qw
& \ctrl{-1}
& \qw
& \qw
& \qw
& \qw
& \qw
& \ctrl{-1}
& \qw
& \qw
& \qw
& \ctrl{-1}
& \qw
& \qw
& \qw
& \targ{}
& \qw
& \qw
\end{quantikz}
}
\caption{Circuit decomposition of the interaction term (pre-optimization). An efficient analytic decomposition uses a Givens rotation of the binary encoding $\ket{010} \leftrightarrow \ket{101}$. The multi-controlled $R_{x}$ rotation can be decomposed into $6$ $CX$ gates using the $ABC$ decomposition \cite{NielsenChuang2010} and by replacing the Toffolis with $RCCX$s. Each $CX$ gate is equivalent to an $RZZ(\pi/2)$ gate up to single-qubit unitaries and a global phase. Thus, the unoptimized decomposition uses at most 10 $RZZ$ gates. }
\label{fig:interaction_RZZ}
\end{center}
\end{figure}

We first consider decomposition of the Hamiltonian. With Trotterization, we can separate the circuits for hopping, electric energy, and the interaction terms. The Pauli decomposition of the local hopping and both electric energy terms over two plaquettes is given by
\begin{equation}
\begin{aligned}
H_{eff, \text{hop, local}} \propto XX + YY \\
H_{eff, \text{$E^{2}$, local}} \supset (I \pm Z)(I \mp Z).
\end{aligned}
\end{equation}
The terms have immediate optimal decompositions with two $RZZ$ gates (this decomposition for hopping is shown in Fig.~\ref{fig:hopping_RZZ}). For the interaction, a better decomposition is a Givens rotation instead of a Pauli decomposition (Fig.~\ref{fig:interaction_RZZ}). 

With these decompositions, we choose a Trotter ordering with minimal circuit depth. The electric energy term is diagonal, so its action may be completely parallelized ($M$). For the hopping and interaction term, since the terms act on adjacent plaquettes, we may apply them in parallel on the even plaquettes ($E$) and the odd plaquettes ($O$). The Hamiltonian evolution can be decomposed into a second-order Trotter formula as 

\begin{equation}
e^{-i H \tau} \approx \prod_{n}^{\tau  / d \tau}
M\!\left(\text{d}\tau / 2\right)
\begin{cases}
E(\text{d}\tau)\,O(\text{d}\tau), & n \ \text{even},\\[2mm]
O(\text{d}\tau)\,E(\text{d}\tau), & n \ \text{odd},
\end{cases}
\,\times\,M\!\left(\text{d}\tau / 2\right),
\end{equation}
where the application of the even and odd unitaries is switched every Trotter step to ensure the Trotterized evolution is order-agnostic.
\subsubsection{State Preparation}
\begin{figure}[!htbp]
\centering 
\includegraphics[width=1.00\textwidth, height=20cm]{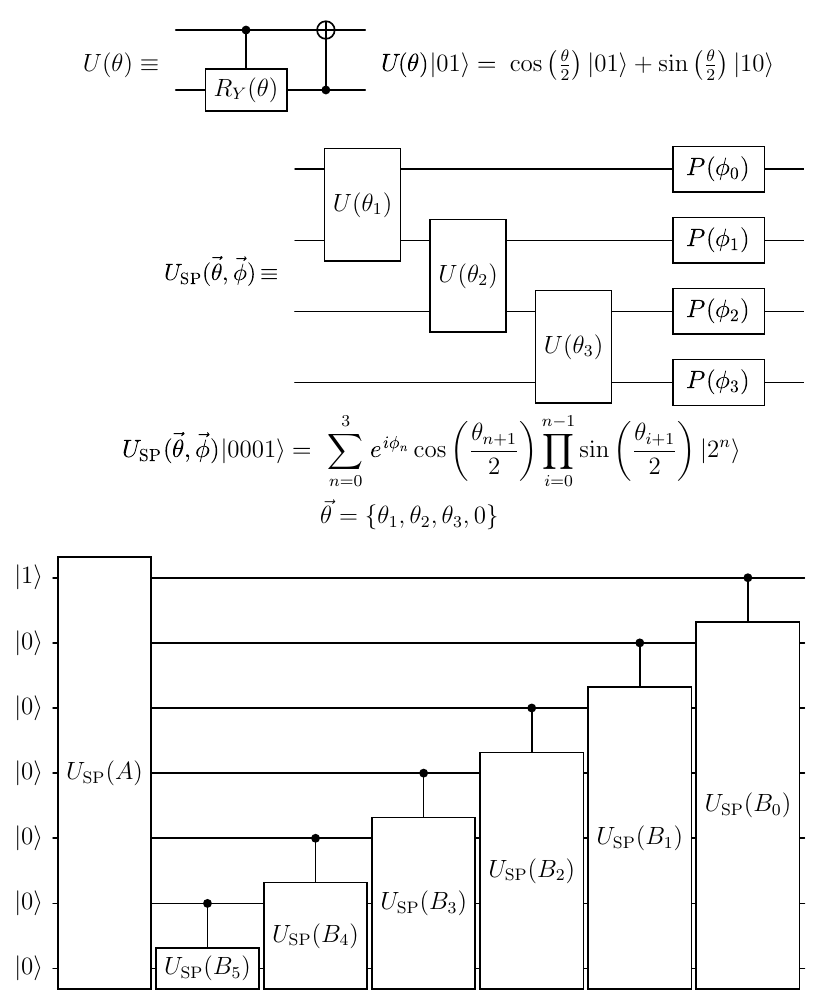}
\caption{
\textbf{Top}: Base primitive for amplitude smearing over
Hamming-weight-one states.
\textbf{Middle}: Recursive use of the primitive for single-particle
state preparation.
\textbf{Bottom}: Recursive construction of arbitrary
two-particle states, with $A(x_1)$ and $B_{x_1}(x_2)$ defined in Eqn. \ref{eqn:A_B}.
}
\label{fig:particle_smearing}
\end{figure}
The single-particle wave-packets in the computational basis are of the following form, with $\sum_n|c_n|^{2}=1$:
\begin{equation}
\begin{aligned}
\ket{\psi(k_0, \sigma_{k}, n_{0})} = \sum_{n=0}^{N-1} c_{n}\!\left(k_0, \sigma_k, n_0\right) e^{i\phi_{n}(n_0)} \ket{2^{2n}}
\end{aligned}
\end{equation}
To create such wave-packets efficiently, we use the linear-depth circuit to construct arbitrary states in the Hamming-weight-one subspace, as described in \cite{farrell2025}. The circuit smears amplitudes through controlled $R_{y}\left( \theta \right)$ rotations, with $\{\theta_{i}\}$ given by $\{c_{n}\}$ and the recursion of the smearing. The construction of the algorithm is presented in Fig.~\ref{fig:particle_smearing}. For constructing two wave-packets on the same chain, if the wave-packets are spatially narrow, then it is sufficient to use the single-packet circuit on separated parts of the chain. 

Finally, to implement step 2 for observing the decay of the scattering state we present a circuit construction for the unary encoding of a generic two-particle Bethe state based on the recursive circuit for a single excitation. We first write the product plane wave in a factorized form 
\begin{equation}
\begin{aligned}
\ket{\psi(k_1, k_2)} = \frac{1}{\mathcal{N}} \left( \sum_{x_2 > x_1 } \left(e^{i k_1 x_1}e^{i k_2 x_2} + S(k_1, k_2) e^{i k_2 x_1}e^{i k_1 x_2} \right) \ket{x_1, x_2}\right)  \\
 =  \frac{1}{\mathcal{N}}\left( \sum_{x_2 > x_1 } \left(e^{i \left(k_1  + k_2\right)x_1} \left(e^{i k_2 (x_2 - x_1)} + S(k_1, k_2) e^{i k_1 (x_2 - x_1) } \right)\right) \ket{x_1, x_2}\right) 
\end{aligned}
\end{equation}

In this form with absolute coordinate $x_1$ and relative coordinate $x_2 - x_1$, the excitations and their phases are essentially decoupled. We may now use the single-particle excitation circuit to initialize the first excitation at coordinate $x_1$ with phase $e^{i (k_1 + k_2) x_1}$, and then condition on the first excitation to create the second excitation with the remaining phase at coordinates $x_2 > x_1$. Explicitly, we can define an $A(x_1)$ function for the first excitation and $B_{x_1}(x_2)$ for the second excitation

\begin{equation}
\begin{aligned}
A(x_1)
&=
\frac{e^{i(k_1+k_2)x_1}}{\mathcal N}
\left[
\sum_{y>x_1}
\left|
e^{ik_2(y-x_1)}
+S(k_1,k_2)e^{ik_1(y-x_1)}
\right|^2
\right]^{1/2},
\\
B_{x_1}(x_2)
&=
\frac{
e^{ik_2(x_2-x_1)}
+S(k_1,k_2)e^{ik_1(x_2-x_1)}
}{
\left[
\sum_{y>x_1}
\left|
e^{ik_2(y-x_1)}
+S(k_1,k_2)e^{ik_1(y-x_1)}
\right|^2
\right]^{1/2}
},
\qquad x_2>x_1,
\\
\text{with}\quad\mathcal N
&=
\left[
\sum_{x_1<x_2}
\left|
e^{ik_1x_1+ik_2x_2}
+S(k_1,k_2)e^{ik_2x_1+ik_1x_2}
\right|^2
\right]^{1/2}.
\label{eqn:A_B}
\end{aligned}
\end{equation}

$A(x_1)$ and $B_{x_1}(x_2)$ are defined such that $\sum_{x_1}|A(x_1)|^2=1$ and
$\sum_{x_2>x_1}|B_{x_1}(x_2)|^2=1$ and such that 

\begin{equation}
\begin{aligned}
A(x_1)B_{x_1}(x_2)
&=
\frac{e^{i(k_1+k_2)x_1}}{\mathcal N}
\left[
e^{ik_2(x_2-x_1)}
+S(k_1,k_2)e^{ik_1(x_2-x_1)}
\right] \\
&=
\frac{1}{\mathcal N}
\left[
e^{ik_1x_1+ik_2x_2}
+S(k_1,k_2)e^{ik_2x_1+ik_1x_2}
\right],
\end{aligned}
\end{equation}

With these distributions, the circuit for the two-particle smearing is given in Fig. \ref{fig:particle_smearing}. Overall, this circuit is not necessarily hardware-efficient, but has the advantage of being transparent to implement and general enough to prepare any two-particle Bethe state. For demonstrating extraction of the resonance on a quantum computer, this circuit suffices. Related exact encoders for arbitrary fixed-Hamming-weight subspaces have also been developed \cite{Farias2025}. 

\subsection{Simulation Results}
Using the  circuit elements described above, we simulate collisions and oscillations on a 14-plaquette chain in the toy model. The encoding uses 28 qubits. The collision simulation was run on the Quantinuum \texttt{H2-2E} emulator with noise modeling and subsequently on  \texttt{H2-2} hardware. 

Given $\alpha_{1}, \alpha_{2},$ and fixed $+$ electric energy, the $\ket{e}$ electric energy is chosen to be near resonance. To observe the resonance and dynamics, we have chosen $\{\alpha_{1}, \alpha_{2}, \epsilon_{+}, \epsilon_{e} \} = \{-1.2, 2, 5.\overline{33}, 10.80 \}$. The numerical details for this model are provided in App.~\ref{app:toy_model}.
\subsubsection{Wave-packet Collision}

\begin{figure}[!htbp]
    \vspace{-1.0cm}
    \centering
    \begin{minipage}[c]{0.43\textwidth}
        \centering
        \includegraphics[width=\linewidth]{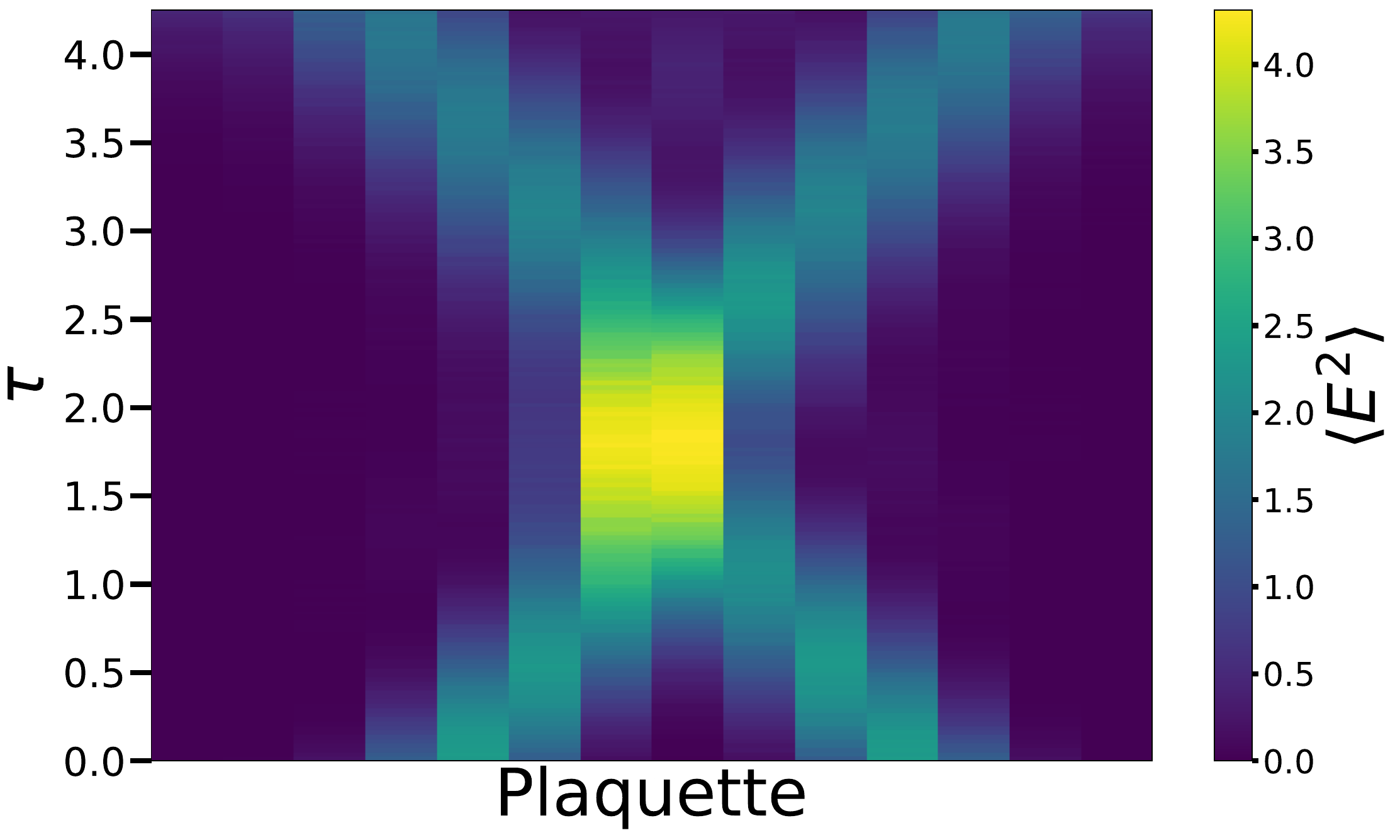}
    \end{minipage}
    \begin{minipage}[c]{0.43\textwidth}
        \centering
        \includegraphics[width=\linewidth]
        {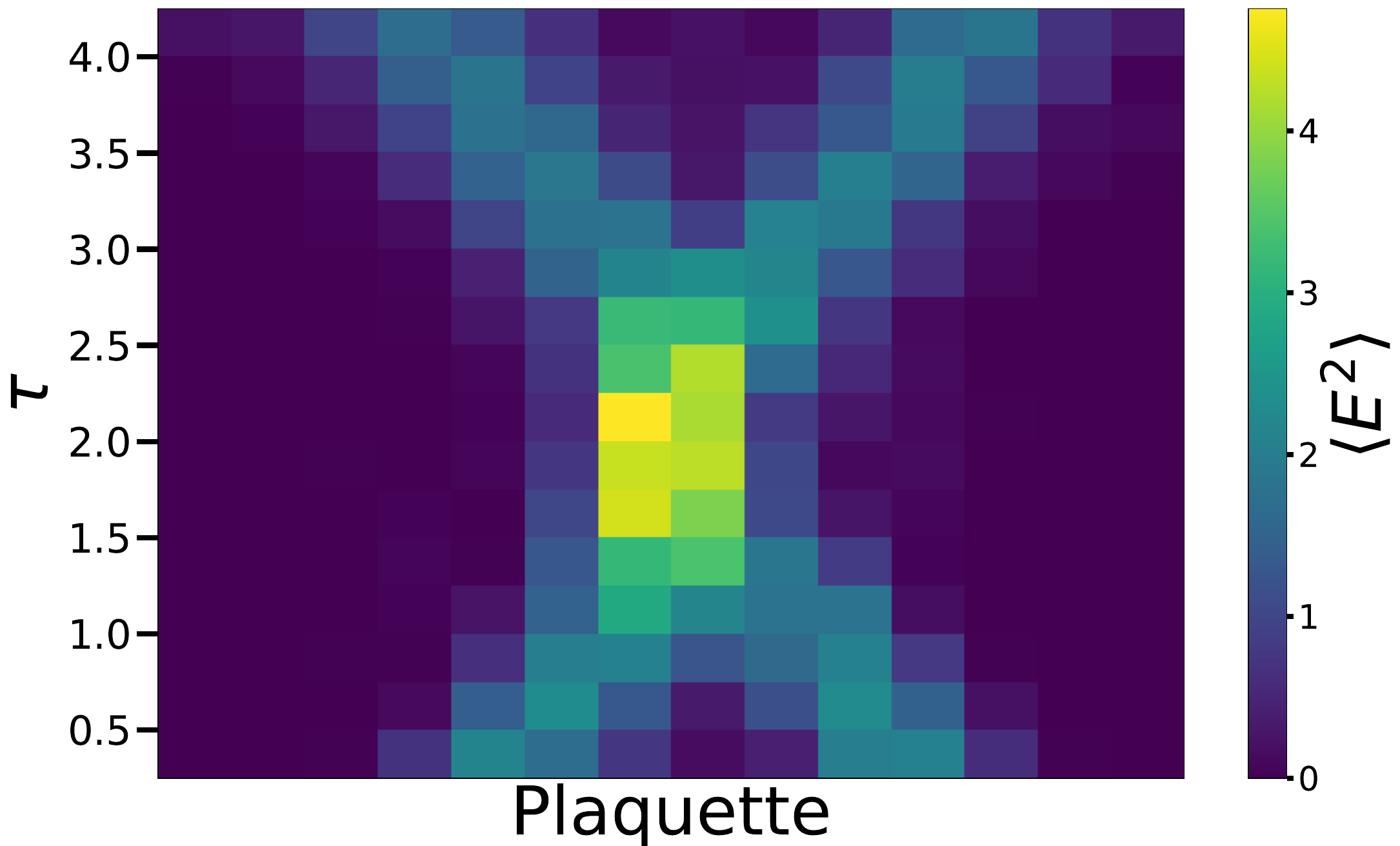}
    \end{minipage}

    \vspace{0.5em}

    \begin{minipage}[c]{0.80\textwidth}
        \centering
        \includegraphics[width=\linewidth]
        {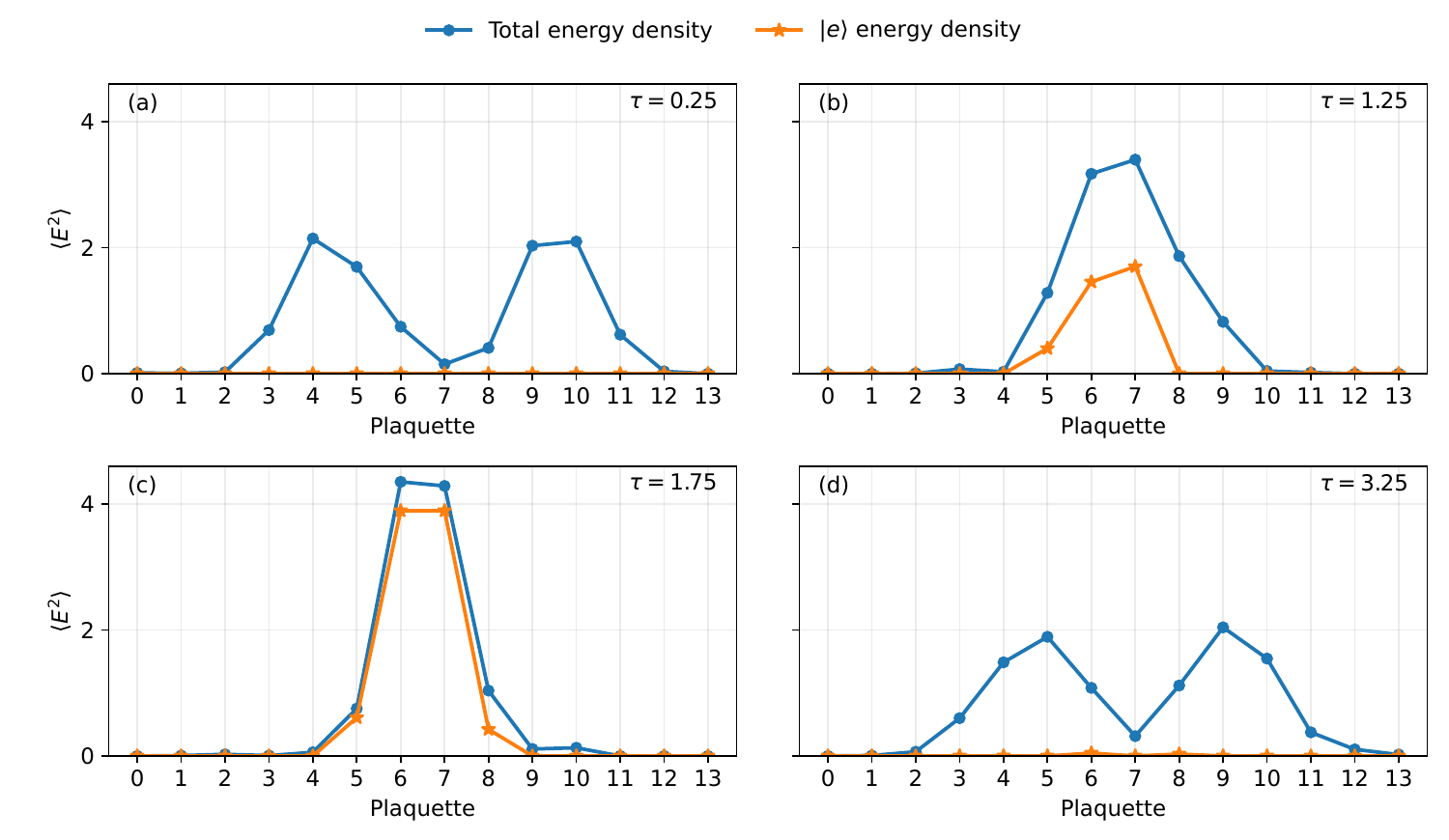}
    \end{minipage}

    \vspace{0.5em}
    
    \begin{minipage}[c]{0.80\textwidth}
        \centering
        \includegraphics[width=\linewidth]
        {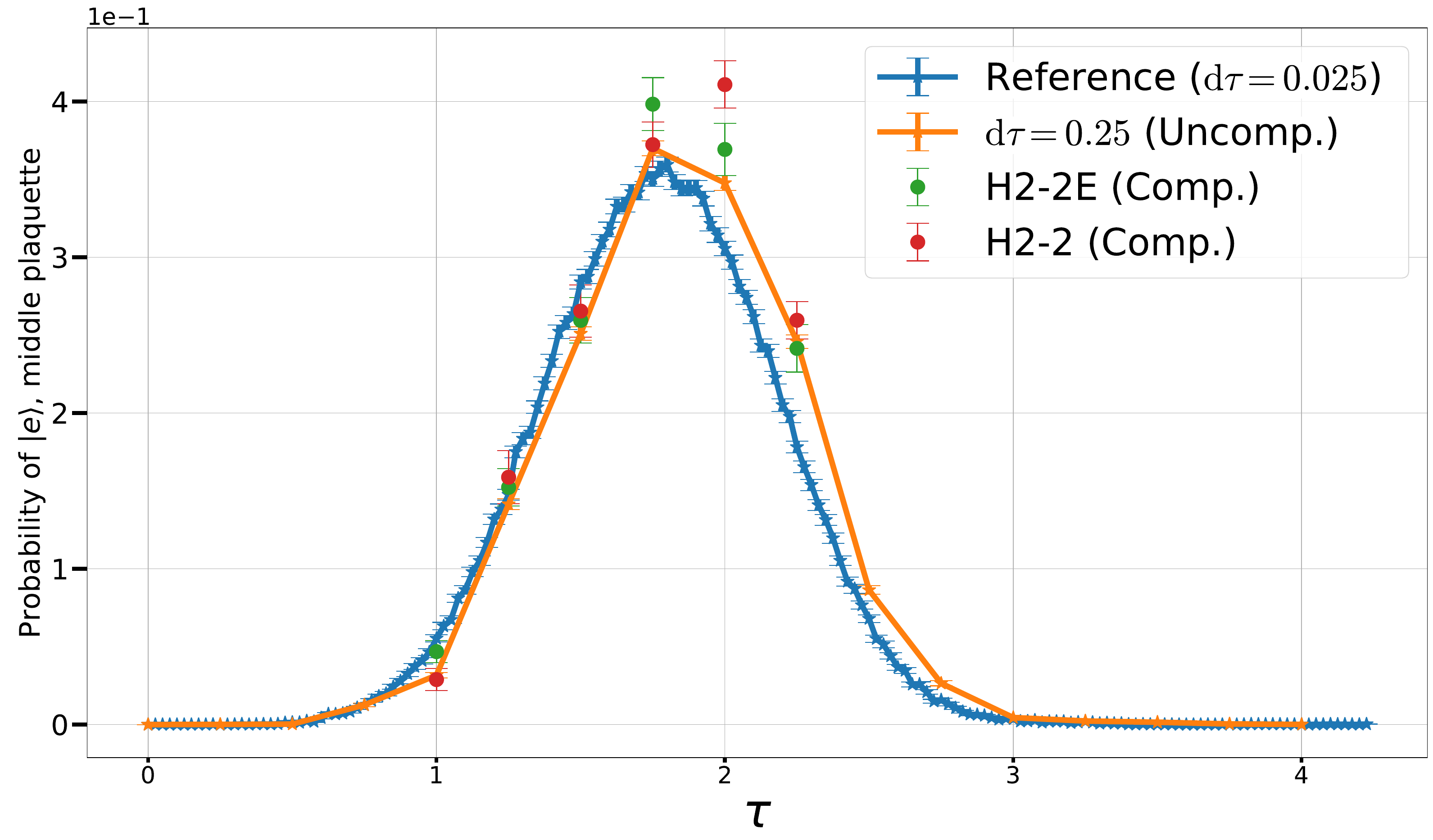}
    \end{minipage}

    \caption{
        \textbf{Top, left:} Collision with $\text{d}\tau=0.025$.
        \textbf{Top, right:} Quantinuum \texttt{H2-2E} emulator collision result using compressed, $\text{d}\tau = 0.25$ circuits ($t = 0$ omitted).
        \textbf{Middle}: Time snapshots of the \texttt{H2-2E} collision simulation. Error bars are omitted for visual clarity. Both $\ket{e}$ energy density and total energy density are tracked, showing how the wave-packets collide and produce the localized $\ket{e}$ as a function of time. 
        \textbf{Bottom:} $\ket{e}$ production on the central plaquette before and after optimization and compression, with points around the peak run on $\texttt{H2-2E}$ and $\texttt{H2-2}$. The error bars on the curves and points are shot noise. As seen, the measured $\ket{e}$ from the Trotterized and compressed circuits run on hardware and emulator generally reproduces the reference results. 
        }
    \label{fig:toy_model_dynamics}
\end{figure}

The baseline near-continuous-time ($\text{d}\tau = 0.025$) wave-packet collision   is shown as a heatmap in the first row of Fig.~\ref{fig:toy_model_dynamics}. For the runs on emulator and hardware, we choose a Trotter step size of $\text{d} \tau = 0.25$ and reduce circuit depth and $RZZ$ gate count through optimization and compression. The depth and $RZZ$ counts scale as $\sim 7n$ and $\sim 25n$, respectively, for the resulting circuits. For error mitigation, dynamical decoupling is turned on and the shot data is projected onto a defined physical plaquette subspace, following the general strategy of symmetry-based error mitigation \cite{BonetMonroig2018}. These details are elaborated on in App. \ref{app:toy_model}. Results on Quantinuum hardware are found to accurately simulate the dynamics of the scattering process. A  clear time-delay signature is evident for the resonant collision, as well as strong production of the resonant $\ket{e}$ state. 

\subsubsection{Scattering State Decay}
\begin{figure}[!htbp]
\begin{center}
\hspace{-1cm}
\includegraphics[width=0.50\textwidth]{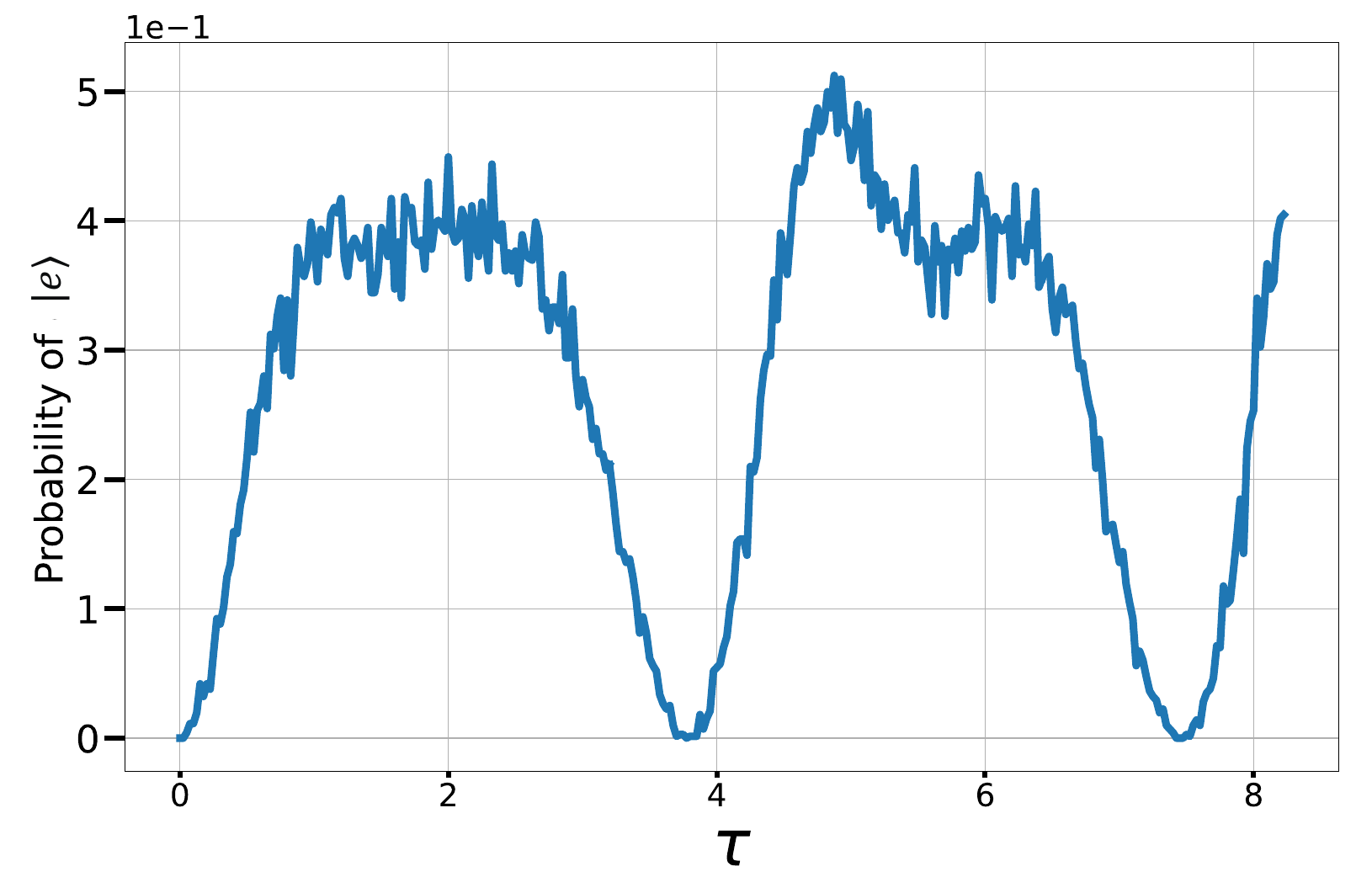} 
\includegraphics[width=0.50\textwidth]{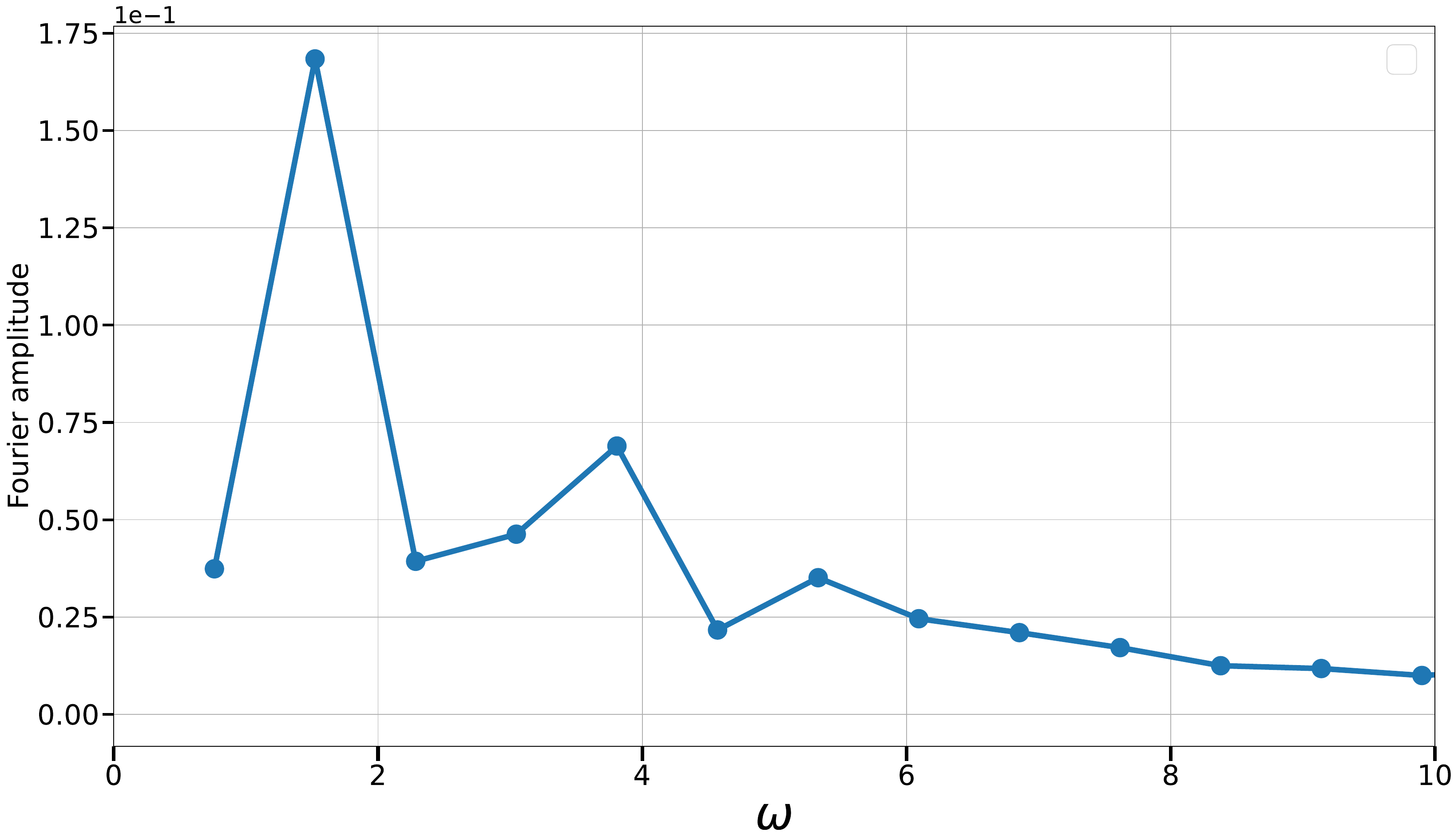} 
\caption{\textbf{Left:} Probability of the $K=0$ $\ket{e}$ state from the non-dressed scattering state. Evolution is Trotterized with d$\tau = 0.025$ to minimize Trotter error, with $10^4$ shots per time point. The minima occur at intervals of roughly $\tau \in [3.7, 3.75]$, which corresponds to the numerical $\Delta_{\text{AC}} \approx 1.69$. \textbf{Right:} Frequency decomposition of the time series data for $\tau = 8.25$ and d$\tau = 0.025$. The $\omega$ with the highest peak is closest to $\Delta_{\text{AC}}$.} 
\label{fig:dt=0.025_oscllication}
\end{center}
\end{figure}

Given the observed time delay, we can isolate the two-particle scattering state $\ket{k, -k}$ on which the Gaussian wave-packet is centered, time evolve this delocalized state, and observe its Rabi oscillation. The $\Delta_{\text{AC}}$ can be measured by doing a frequency decomposition of the evolution time-series data, with energy resolution given by $2 \pi / \tau$ for evolution up to $\tau$. For optimal extraction, a low sampling rate of $N$ points over a large $\tau$ could suffice to give an accurate measurement of $\Delta_{\text{AC}}$. However, for a Trotterized evolution circuit, the Trotter error per point scales as a function of $\tau / N$. This consideration, along with additional information, like the density of the spectrum and general location of $\Delta_{\text{AC}}$, can inform the optimal $\tau$ and $N$ for which one can recover the avoided crossing gap energy. For the purposes of demonstrating extraction, we forgo this analysis. The scattering state decay circuit is simulated only and not run on emulator or hardware.

In this restricted model, the scattering states only interact with the single bound state, so the $\Delta_{\text{AC}}$ responsible for the resonance is read out by examining the probability of measuring the $K=0$ bound state as a function of time. If the Hamiltonian is well approximated by the $2 \times 2$ model, the probability of overlap between the undressed scattering state and bound state should have the largest contribution from the function $\sin^{2}{\left(\frac{\Delta_{\text{AC}}}{2} t \right)}$. In a frequency decomposition of the time-series data, the largest peak should then correspond to the binned $\omega \sim \Delta_{\text{AC}}$. Fig.~\ref{fig:dt=0.025_oscllication} shows this oscillation up to $\tau=8.25$ and its frequency decomposition. The finite evolution window limits the frequency resolution of this demonstration. For $\tau=8.25$, the Fourier-bin spacing is $\Delta\omega=2\pi/\tau_{\max}\approx0.76$.
The observed peak is therefore consistent with the expected
avoided-crossing gap, $\Delta_{\mathrm{AC}}\approx1.70$.

\section{Conclusion}
We conclude by summarizing the main findings of this work, and some directions for future work.  We performed classical real-time simulations of scattering in SU(3) lattice gauge theory on the line-of-plaquettes lattice  including dynamical adjoint excitations. For a range of intermediate couplings, we observed resonant production of heavier ``glueball" states in the collision of two light glueballs. Introducing time-domain observables probing microscopic features of resonance production, we developed procedures for extracting spectral features from quantum simulations. Using an effective model for the resonance, we constructed quantum circuits and obtained high-accuracy simulation results for wavepacket collisions on Quantinuum \texttt{H2-2} hardware. 

The long-term goal of this field is to simulate real-time scattering and other nonequilibrium phenomena in lattice QCD near the continuum limit. Many steps remain before this target is within reach, but great progress has been made over the past decade. There are several natural directions for extension of the present work. Observing  richer collision phenomena expected to emerge at weak coupling  will require higher truncations and efficient simulations, both classical and quantum. The most straightforward way to reduce the number of states and matrix elements is by diagnosing what states are important for the low-energy physics; this requires a better understanding of how the states are dressed as the coupling decreases. On the quantum side, determining the optimal two-qubit gate costs and depth for wave-packet and product preparation would be useful to determine how resource costs scale for Trotterized simulation of both link-local and plaquette-local encodings.

\section{Acknowledgments}
We thank David Stephen and Etienne Granet for helpful comments on the manuscript.
We acknowledge access to quantum computing resources through the
Quantum Computing User Program (QCUP) at Oak Ridge National
Laboratory. These resources were provided by the Oak Ridge
Leadership Computing Facility, a U.S. Department of Energy
Office of Science user facility supported under Contract DE-AC05-00OR22725.

\bibliographystyle{utphys}

\bibliography{refs}
\appendix
\appendix

\section{Derivation of Effective Hopping and Plane Waves for $+$ Through Perturbation Theory}
\label{app:effective}
Second-order perturbation theory tells us that the effective interaction $M$ between the degenerate -1 states is given as 
\begin{equation}
\begin{aligned}
\label{second-deg}
\braket{+_{i}|\lambda^{2} M|+_{j}} = \lambda^{2} \left(\sum_{p} \sum_{P} \sum_{Q} \frac{\left(\Box_{P} + \Box_{P}^{\dagger}\right)_{ip}\left(\Box_{Q} + \Box_{Q}^{\dagger}\right)_{pj}}{E^{0}_{i} - E^{0}_{p}}\right)
\end{aligned}
\end{equation}
\begin{figure}[t]
    \vspace*{-0.50cm}
    \centering
    \includegraphics[width=1.00\textwidth, height=4cm]{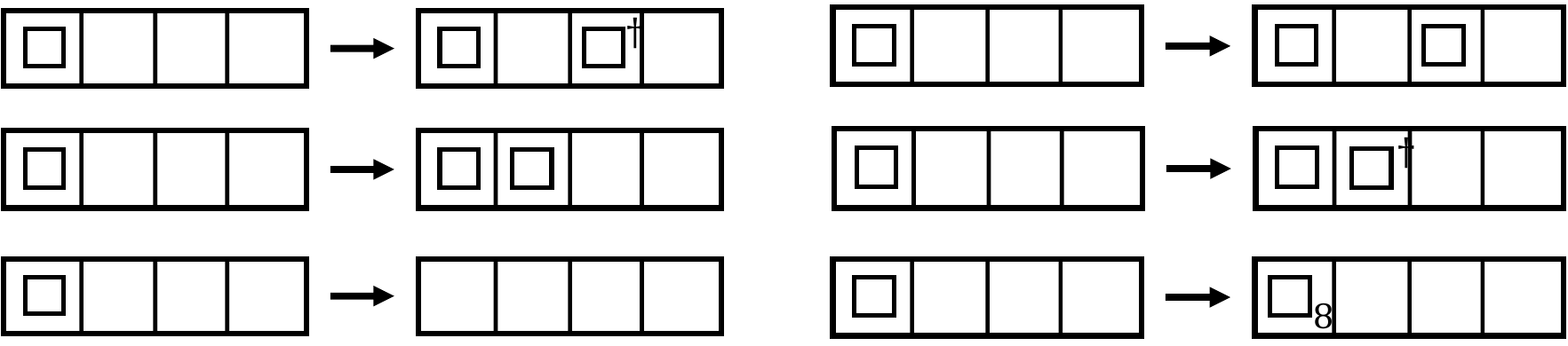}
    \caption{Types of intermediate states that $\ket{\Box}$ transitions to at second order. In the example, the initial state is the box excitation on the first plaquette that transitions to an intermediate state after $\sum_{P} \left(\Box_{P} + \Box^{\dagger}_{P}\right)$ is applied. The $\mathbf{8}$ subscript refers to the plaquette state with $\mathbf{8}$ on all links. The adjacent excitations of the plaquette operator generate superpositions of states with lower-order irreps $(\mathbf{1}, \mathbf{\overline{3}})$ and higher-order irreps $(\mathbf{8}, \mathbf{\overline{6}})$ on the shared middle link.}
    \label{fig:placeholder}
\end{figure}
\begin{figure}[t]
\begin{center}
\hspace*{-1cm}
\includegraphics[width=0.50\textwidth, height=5cm]{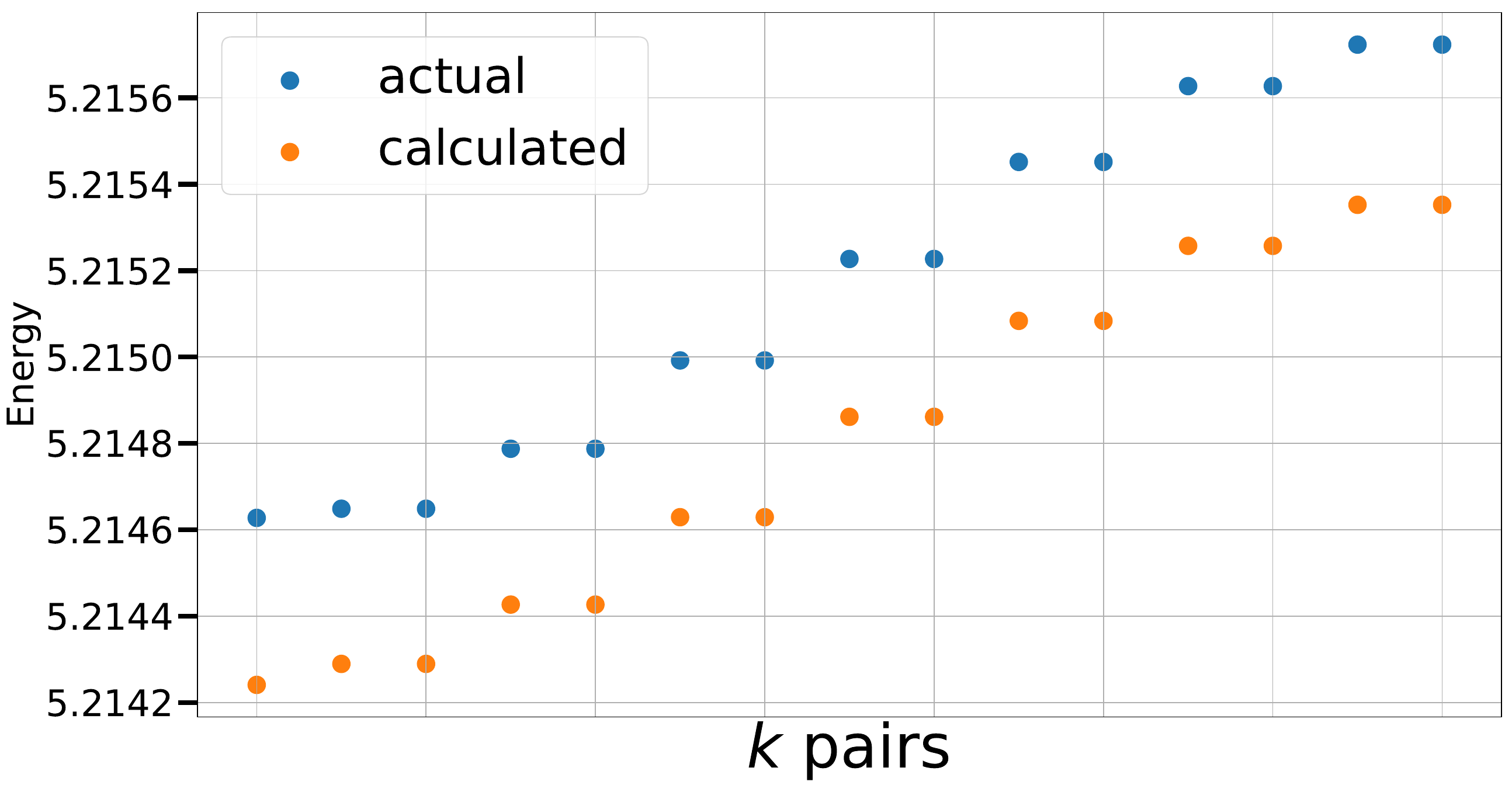} 
\hfill
\includegraphics[width=0.50\textwidth, height=5cm]{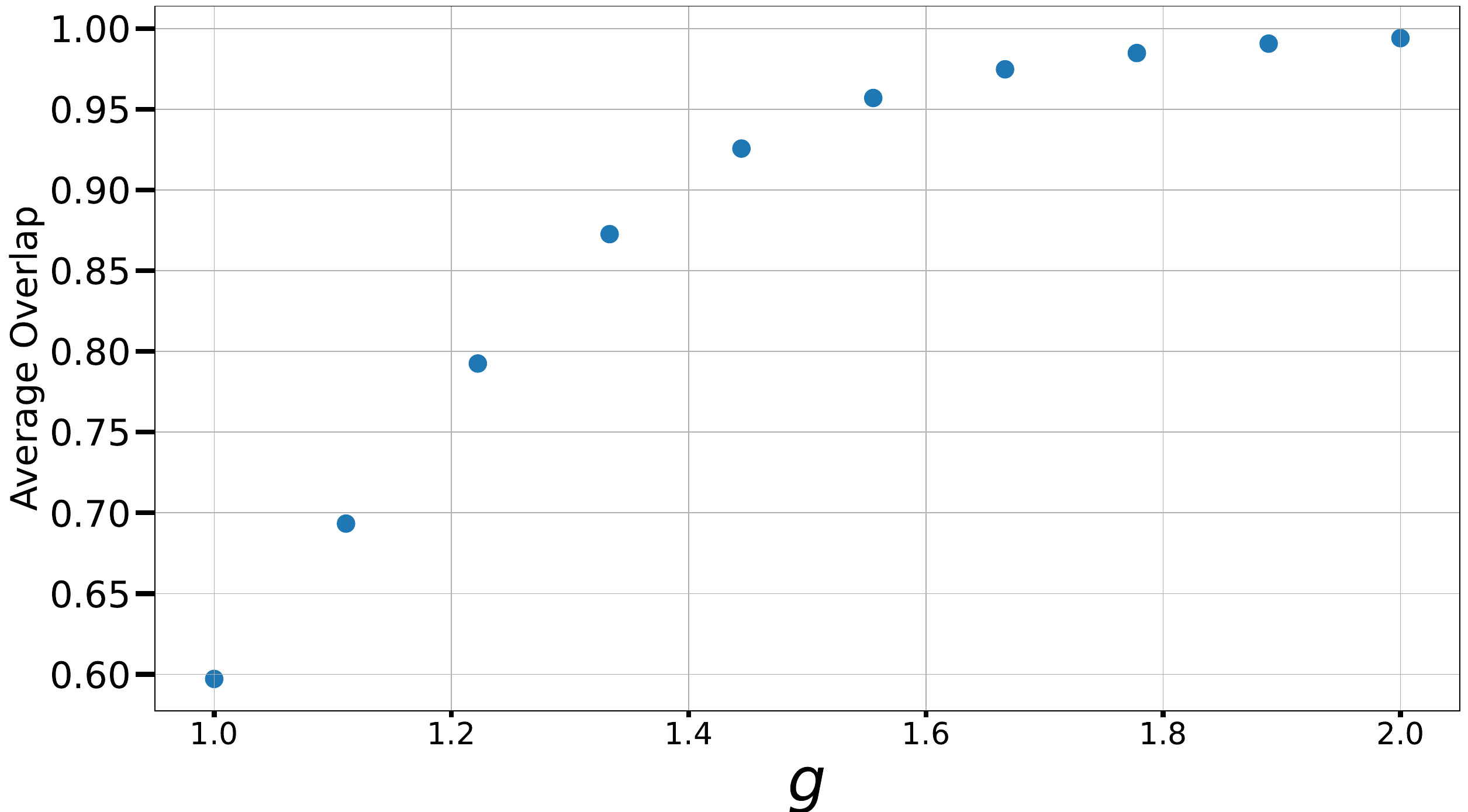}
    \caption{\textbf{Left}: Second-order eigenvalues compared with actual eigenvalues (ordered by increasing momenta) of the rescaled Hamiltonian at $g=2.0, N=15$. With $\lambda = 2/g^{4}$, $\lambda^{3} \sim 10^{-3}$. \textbf{Right}: The average overlap $\braket{k|\tilde{k}}$ between the perturbative eigenvectors $\ket{k}$ and actual eigenvectors $\ket{\tilde{k}}$ for $k \neq 0$ as a function of $g$ with $N = 12$.}
    \label{fig:eigs}
\end{center}
\end{figure}

\noindent where the sum $p$ is a sum over states outside the degenerate box excitation space. Under the magnetic transitions and irrep truncation $\{\mathbf{1}, \mathbf{3}, \mathbf{\overline{3}}, \mathbf{8} \}$, $\{\ket{\Box}_{n}\}$ mixes with 6 different states, as seen in Fig.~\ref{fig:placeholder}. The $\{\ket{\Box^{\dagger}_{x}}\}$ states mix similarly. With these transition amplitudes and energies, the matrix elements of $M$ are
\begin{equation}
\begin{aligned}
&\braket{{i}|\lambda^{2} M|{i}} = -\lambda^{2}\left(\frac{997}{2040} + \frac{3}{8}(N-3)\right) \\
&\braket{i|\lambda^{2} M|i \pm 1} =\frac{19\lambda^{2}}{204}
\\
&\braket{i|\lambda^{2} M|j} = 0 \quad (\text{$j$ otherwise})
\end{aligned}
\end{equation}
The $M$ interaction allows an $\ket{+}$ excitation on plaquette $i$ to go to plaquette $i \pm 1$ (``hop''), so its eigenvectors and eigenvalues capture the effective states for wave-packet propagation in the full theory. The diagonal elements grow with $N$, but the ground state receives the same -$\lambda^{2} (3/8)N$ correction from mixing with excited $\ket{\Box_{n}}$ and $\ket{\Box^{\dagger}_{n}}$ states. Subtracting this energy from the diagonal gives 
\begin{equation}
\begin{aligned}
&\braket{+_{i}|\lambda^{2} M|+_{i}} = \frac{649 \lambda^{2}}{1020}
\end{aligned}
\end{equation}
The matrix $M$ is circulant: each row is the previous row shifted one to the right. Circulant matrices have powers of the roots of unity as amplitudes for their eigenvectors, which are our phase factors for the plane waves, and admit closed-form eigenvalues. For a matrix $\tilde{M}$ of this form, the eigenvectors and eigenvalues are 
\begin{equation}
\begin{aligned}
\ket{k} = \frac{1}{\sqrt{N}}\sum_{n=0}^{N-1}e^{- \frac{2 \pi i n_{k} n}{N}} \ket{n} \\
\epsilon_{k} = \braket{0|\tilde{M}|0} + 2 \braket{0|\tilde{M}|1} \cos\left(\frac{2 \pi n_{k}}{N} \right)
\end{aligned}
\end{equation}
The eigenvalues for $\lambda^{2} M$ are
\begin{equation}
\begin{aligned}
\lambda^{2} E^{2}_{k} = \frac{\lambda^{2}}{2040}\left(1298 + 380\cos\left(\frac{2 \pi n_{k}}{N}\right)\right)
\end{aligned}
\end{equation}

For the effective Hamiltonian in our space
\begin{equation}
\begin{aligned}
H_{eff,+} = \sum_{l} E^{l, 2} - \lambda \, I + \lambda^{2} M 
\end{aligned}
\end{equation}
the dispersion relation for the scaled Hamiltonian is 
\begin{equation}
\begin{aligned}
\label{energy_minus}
\omega_{k,+} = E^{0}_{k} + \lambda E_{k} + \lambda^{2} E^{2}_{k} = \frac{16}{3} - \lambda + \frac{\lambda^{2}}{2040}\left(1298 + 380\cos\left(\frac{2 \pi n_{k}}{N}\right)\right)
\end{aligned}
\end{equation}
Fig.~\ref{fig:eigs} shows agreement between the dispersion relation and eigenvalues of the Hamiltonian at $g = 2.0$, as well as the accuracy of the plane wave approximation to the actual eigenvectors as a function of $g$.

\section{Details of Classical Simulation Methods and Construction}
\label{app:global}
\subsection{Global Basis}
For the truncations and lattices considered, we use a global basis to simulate evolution on the lattice rather than a local one. In a global basis, instead of considering the lattice Hilbert space as a tensor product of the individual link and vertex Hilbert spaces, the Hilbert space contains only the gauge-invariant configurations on the lattice. The resulting Hilbert space is significantly smaller, at the expense of a larger and denser $\Box$ operator. While the local encoding remains ideal for quantum simulations, the global encoding is preferable for classical simulations, where space is expensive but operations are cheap. An additional benefit is that features of the Hamiltonian, like the spectrum, are much easier to analyze in the global encoding, especially when global symmetries create superselection sectors. 
\subsubsection{Methods for Constructing the Global Basis Hilbert Space}
The gauge-invariant states of the lattice must be first enumerated in order to construct the Hamiltonian acting on the physical Hilbert space of the theory. There are two methods for this enumeration:

\begin{enumerate}
    \item Starting from an initial gauge-invariant state (like the vacuum), repeatedly apply the local plaquette operator $(\Box + \Box^{\dagger})$ (restricted to some local irrep truncation) to reach other states. This is akin to a \textbf{Breadth-First-Search (BFS)}, where states are discovered ``expanding'' outward from a seed state
    \item Stitch together link and vertex combinations that satisfy the local Gauss's law. This is akin to a \textbf{Depth-First-Search (DFS)}, where the states are discovered branch by branch
\end{enumerate}

\begin{figure}[H]
    \centering

    \makebox[\textwidth][c]{%
        \includegraphics[width=0.78\textwidth]
        {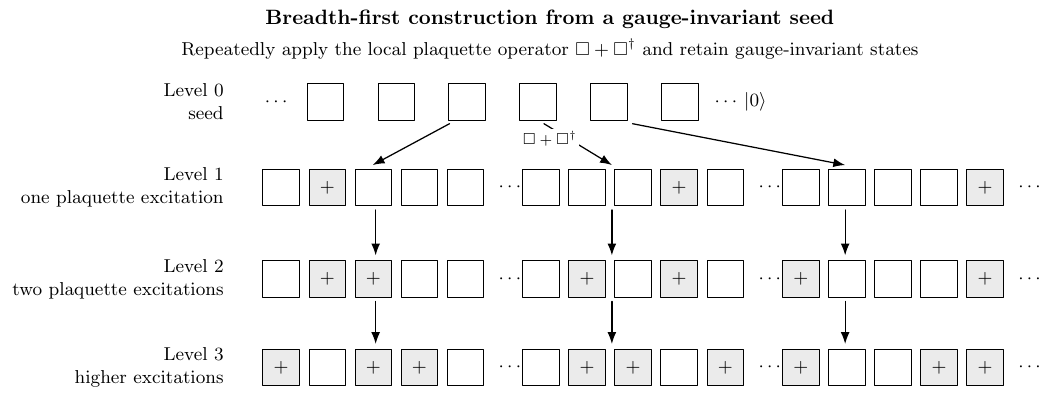}
    }

    \vspace{0.15cm}

    \makebox[\textwidth][c]{%
        \hspace{0.6cm}
        \includegraphics[width=0.78\textwidth]
        {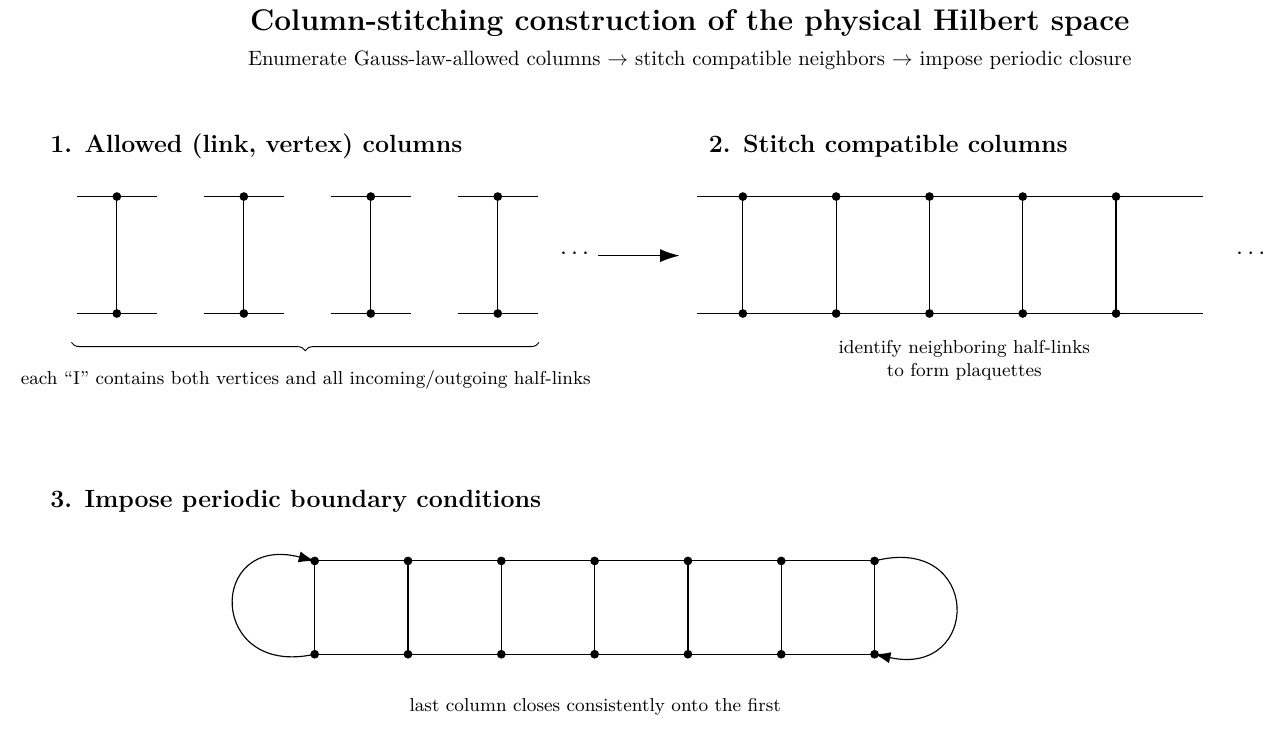}
    }

    \caption{Enumeration methods for the global gauge-invariant Hilbert space
    for the chain of plaquettes with PBCs. The top diagram corresponds to the
    ``BFS'' method and the bottom diagram corresponds to the ``DFS'' method.}
    \label{fig:enumeration_methods}
\end{figure}

\begin{figure}[H]
\centering
\includegraphics[width=0.50\textwidth, height=5cm]{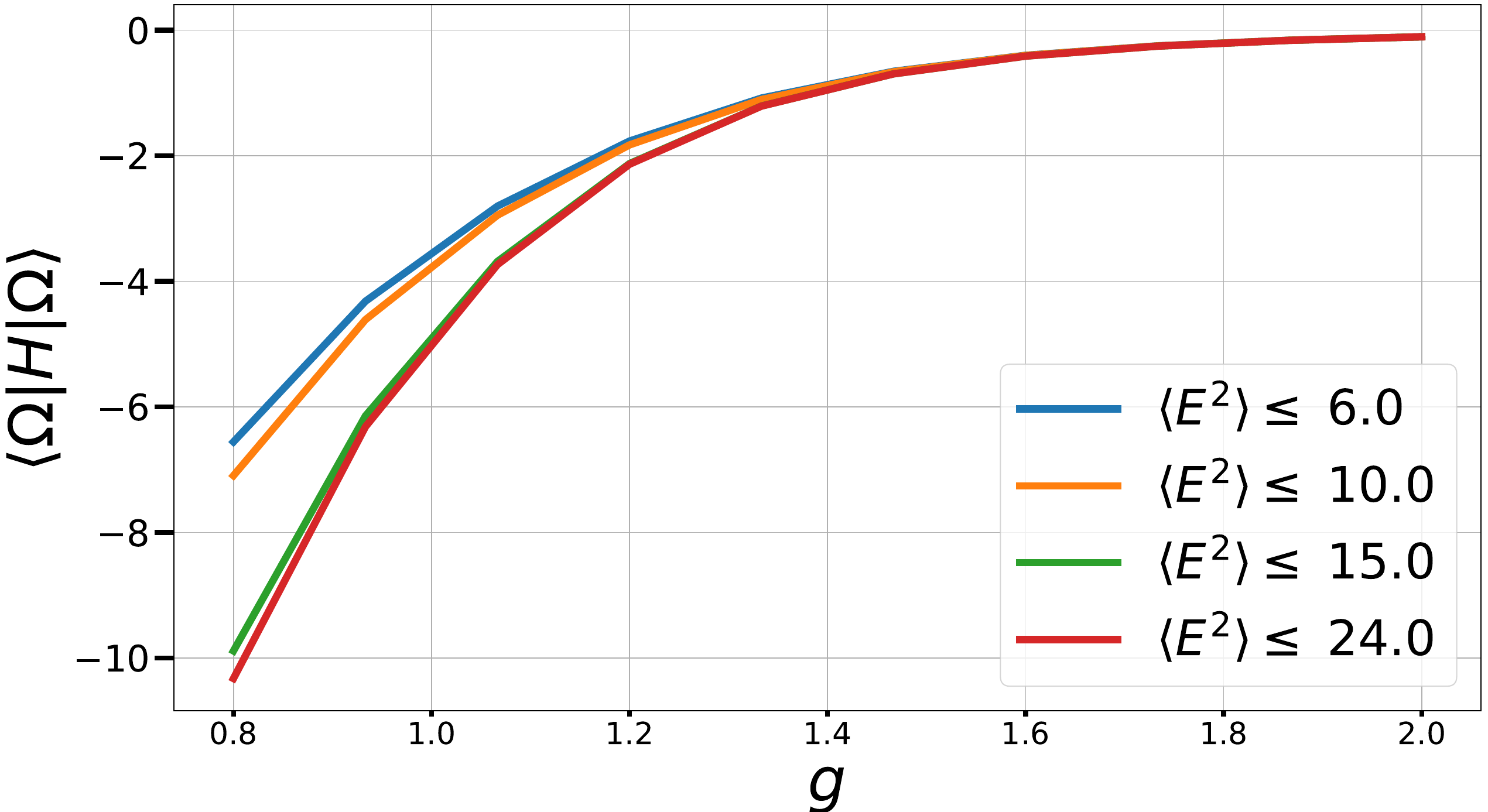}
\caption{The ground state energy as a function of the coupling $g$ for $d = 2, 3 \times 3, T1$, with different global energy truncations. The total Hilbert space is enumerated with $\braket{E^2} \leq 24$; $\braket{E^2} \leq 15$ is sufficient to get the ground state for $g \sim 0.8$.}
\label{fig:3x3_gs}
\end{figure}

The two methods are illustrated in Fig.~\ref{fig:enumeration_methods}. For $d = 3/2$, the DFS stitching method is significantly faster than the BFS method because the stitching is linear and the number of local link/vertex combinations is constrained. In higher dimensions $d$, both these advantages become disadvantages and the BFS method becomes the better method. One caveat of the BFS method is that the number of local excitations needed to reach all states is \textit{a priori} unknown. In most situations, the entire physical Hilbert space given a $B$ truncation, irrep truncation, and coupling $g$ is unnecessary, however. An example is presented in Fig.~\ref{fig:3x3_gs}, where only a tenth of the total physical Hilbert space is necessary to describe the ground state at weak coupling ($g \sim 0.8$). 

\subsubsection{Hilbert Space for $N=12$ and $N=16$}

\begin{table}[H]
    \centering
    \begin{tabular}{c|cc}
        \hline
        $N$, Truncation & \# of States & \# of Magnetic Transitions \\
        \hline
        12, $\braket{E^{2}} \leq 24$ & $8 \times 10^{4}$ & $1.1 \times 10^{6}$  \\
        16, $\braket{E^{2}} \leq 28$ & $1.4 \times 10^{6}$ &  $1.2 \times 10^{7}$ \\
        \hline
    \end{tabular}
    \caption{Size of the gauge-invariant Hilbert space and number of nonzero Hamiltonian matrix elements for the $N=12$ and $N=16$ plaquette chains.}
    \label{tab:hilbert_space_sizes}
\end{table}

Table~\ref{tab:hilbert_space_sizes} shows the number of states and magnetic transitions for the two chains considered with global electric energy truncations. The truncations were chosen to simulate the low-energy dynamics accurately at the intermediate- to strong-coupling regime and observe the transition between the scattering state and $\Box_{3, \times 2, 3}$. For $g \sim 1.3$, where the transition occurs, the dressed Hamiltonian eigenstates, especially the vacuum, become delocalized and mix with all energy sectors. With larger $N$, a larger truncation is required to faithfully represent dressed states and their transitions. 

\section{Derivation, Analysis, and Hardware Optimization of the Toy Model}
\label{app:toy_model}
\subsection{Derivation}
We first motivate the toy model with general perturbative arguments for states embedded in a much larger theory. We consider a translationally invariant Hamiltonian on a lattice with $N$ sites where a characteristic bound state (like the triplet molecule) is embedded in the two-particle scattering spectrum. Let $H = H_{0} + \lambda V$, where $H_{0}$ contains the diagonal energies $\{ E_{i}\}$ and $V$ is a potential. Define the $\left(\binom{N}{2} + N\right)$-dimensional Hilbert space as $\mathcal{H}_{eff}$ and the complementary Hilbert space as $\mathcal{H}_{\perp}$. Without loss of generality, we work in a basis that diagonalizes the first-order perturbation
within $\mathcal{H}_{eff}$, and absorb
the resulting first-order energy shifts into the diagonal energies. With that, we define the Hamiltonian in the Schrieffer-Wolff basis \cite{SchriefferWolff1966,Bravyi2011SW}. The Schrieffer-Wolff transformation $e^{S}$ for $S = \lambda V_{ij} / (E_{i} - E_{j})$ gives us 
\begin{equation}
\begin{aligned}
&H' = e^{S}He^{-S} = e^{S}\left(H_{0} + \lambda V \right)e^{-S} = H_{0} + \lambda V + [S, H_{0}] + [S, \lambda V] + \frac{1}{2}[S, [S, H_0]] + O(\lambda^{3}) \\
&(H')_{ij} = (H_{0})_{ij} + \frac{\lambda^{2}}{2}\sum_{k} \left(\frac{V_{ik}V_{kj}}{E_{i}-E_{k}} +  \frac{V_{ik}V_{kj}}{E_{j}-E_{k}} \right) + O(\lambda^{3}) \approx (H_{0})_{ij} + V'_{ij}
\end{aligned}
\end{equation}
To leading order, the transition between $\ket{i} \in \mathcal{H}'_{eff}$ and $\ket{j} \in \mathcal{H}'_{\perp}$ is given by 
\begin{equation}
\begin{aligned}
\left| \braket{i| e^{-i H' t} | j}\right|^{2} \propto \frac{4V^{'2}_{ij}}{4V^{'2}_{ij} + (E'_i - E'_j)^{2}} \sin^{2} \left(\frac{\sqrt{4V^{'2}_{ij} + (E'_i - E'_j)^{2}}}{2} t \right)
\end{aligned}
\end{equation}
For $V'_{ij} \ll |E'_{i} - E'_{j}|$, the transition amplitude is suppressed quadratically and evolution within $\mathcal{H}'_{eff}$ stays within $\mathcal{H}'_{eff}$. Now, for evolution within $\mathcal{H}'_{eff}$, if the potential is perturbative $(\lambda \ll 1)$, then the leading-order interaction is the second-order interaction with two applications of $V$. If $V$ is an annihilation/creation operator at a site, then the effective potential just provides a hopping term and an interaction term between the position basis states and the bound state. 
 
These arguments motivate the toy model as a representative of the resonance found in the $SU(3)$ Kogut-Susskind Hamiltonian at strong coupling, where the potential is naturally perturbative. The resonance observed in the full model, however, occurs at $g \sim 1.3$, where these perturbative arguments break down. This toy model would be fully realizable in the full theory if the resonance found was at coupling $g > 1.7$.
 
\subsection{Numerical Analysis}
For $\{\alpha_{1}, \alpha_{2}, \epsilon_{+}, \epsilon_{e} \} = \{-1.2, 2, 5.\overline{33}, 10.80 \}$, we may find the spectrum and dressed states directly. For $N=14$ in the $K=0$ sector, the physical Hilbert space just contains 8 states: the single $\ket{e}$ bound state and the seven scattering states $\ket{k, -k}$. The hard-core conditions on the $+$ excitations mean that the scattering states are defined with the Bethe ansatz. For the two-particle wavefunction, $x_1 \neq x_2$ means that the scattering phase $S(k_1, k_2) = -1$. The states in the $K=0$ sector are 
\begin{equation}
\begin{aligned}
\ket{\psi_{k}(x_1, x_2)} &\propto \sum_{x_1}\sum_{x_2 > x_1} \left(e^{-i k \left(x_2 - x_1\right)} - e^{-i k \left(x_1 - x_2\right)}\right) \ket{x_1, x_2} \\ &\propto i\sum_{x_{1}}\sum_{x_2 > x_1}\sin \left( k(x_2 - x_1)\right) \ket{x_1, x_2} \\ &\propto \sqrt{2}  \sum_{r=1}^{6} \sin \left( k r \right) \ket{r} + \sin \left( 7k \right) \ket{7}
\end{aligned}
\end{equation}
where the last line comes from defining an orthogonal set of $\ket{r}$ states as 
\begin{equation}
\begin{aligned}
&\ket{r} = \frac{1}{\sqrt{14}} \sum_{i} \ket{i, i + r} \quad \text{for $r \in [1, 6]$} \\
&\ket{7} = \frac{1}{\sqrt{7}} \sum_{i}^{6} \ket{i, i + 7}
\end{aligned}
\end{equation}
with normalization differing on the $r=7$ state due to the half-orbit. With periodic boundary conditions, the quantization condition is that 
\begin{equation}
\begin{aligned}
e^{-i k N}S(k_1, k_2) = 1 \implies \\
kN + (2m + 1)\pi = 2n\pi \implies k = \frac{\left(2m + 1 
\right)\pi}{N}
\end{aligned}
\end{equation}
With this quantization and in this basis, the $\{\ket{k}\}$ have dispersion $\omega_{k} = 2\epsilon_{+} + 4\alpha_{1} \cos \left( k \right)$. The interaction with the $K=0$ $\ket{e}$ state becomes
\begin{equation}
\begin{aligned}
\ket{e, K=0} = \frac{1}{\sqrt{14}} \sum_{i} \ket{e}_{i} \quad  \braket{e, K=0 | H_{int} | k} \propto \alpha_{2}  \sin(k)
\end{aligned}
\end{equation}
Introducing the proper normalizations, the numerical $8 
\times 8$ matrix with the basis $\{\ket{e}, \ket{k}\}$ is

\[
\hspace{-1cm}
H_{K=0}
=
\begin{pmatrix}
10.8000
& 0.2379
& 0.6665
& 0.9632
& 1.0690
& 0.9632
& 0.6665
& 0.2379
\\[2pt]
0.2379
& 5.9870
& 0
& 0
& 0
& 0
& 0
& 0
\\[2pt]
0.6665
& 0
& 6.9139
& 0
& 0
& 0
& 0
& 0
\\[2pt]
0.9632
& 0
& 0
& 8.5840
& 0
& 0
& 0
& 0
\\[2pt]
1.0690
& 0
& 0
& 0
& 10.6667
& 0
& 0
& 0
\\[2pt]
0.9632
& 0
& 0
& 0
& 0
& 12.7493
& 0
& 0
\\[2pt]
0.6665
& 0
& 0
& 0
& 0
& 0
& 14.4195
& 0
\\[2pt]
0.2379
& 0
& 0
& 0
& 0
& 0
& 0
& 15.3463
\end{pmatrix}.
\]

Since the off-diagonal terms in the matrix are smaller than the average energy gap, it is expected that we can find the adiabatic scattering state responsible for the resonance simply by diagonalizing the matrix and ordering the dressed states based on their overlap with the bare states. For interaction terms comparable with the bare masses, the dressed energies and states from the bare states can be found through an iterative diagonalization algorithm, like the Jacobi algorithm. Diagonalizing the matrix and then picking an ordered basis based on overlap with the undressed states, the shifted energies are 

\[
\hspace{-1cm}
H_{\mathrm{diag}}
=
\begin{pmatrix}
11.5558 & 0 & 0 & 0 & 0 & 0 & 0 & 0 \\
0 & 5.9711 & 0 & 0 & 0 & 0 & 0 & 0 \\
0 & 0 & 6.7695 & 0 & 0 & 0 & 0 & 0 \\
0 & 0 & 0 & 8.1742 & 0 & 0 & 0 & 0 \\
0 & 0 & 0 & 0 & 9.8562 & 0 & 0 & 0 \\
0 & 0 & 0 & 0 & 0 & 13.1997 & 0 & 0 \\
0 & 0 & 0 & 0 & 0 & 0 & 14.5764 & 0 \\
0 & 0 & 0 & 0 & 0 & 0 & 0 & 15.3634
\end{pmatrix}.
\]

Since the $k = \pi /2$ state has bare energy closest to the bound state, it resonates most strongly. The eigenvector content for the $\ket{E_{+}}$ state at energy $\approx 11.55$ and the $\ket{E_{-}}$ state at energy $\approx 9.85$ is given by 

\begin{equation}
\begin{aligned}
|E_{-}\rangle
={}&
-0.53434\,|e_0\rangle
-0.03285
\left|
\frac{\pi}{14},-\frac{\pi}{14}
\right\rangle_{\text{bare}}
\\
&-0.12104
\left|
\frac{3\pi}{14},-\frac{3\pi}{14}
\right\rangle_{\text{bare}}
-0.40453
\left|
\frac{5\pi}{14},-\frac{5\pi}{14}
\right\rangle_{\text{bare}}
\\
&+0.70486
\left|
\frac{\pi}{2},-\frac{\pi}{2}
\right\rangle_{\text{bare}}
+0.17789
\left|
\frac{9\pi}{14},-\frac{9\pi}{14}
\right\rangle_{\text{bare}}
\\
&+0.07805
\left|
\frac{11\pi}{14},-\frac{11\pi}{14}
\right\rangle_{\text{bare}}
+0.02315
\left|
\frac{13\pi}{14},-\frac{13\pi}{14}
\right\rangle_{\text{bare}},
\end{aligned}
\end{equation}

\begin{equation}
\begin{aligned}
|E_{+}\rangle
={}&
0.55196\,|e_0\rangle
+0.02358
\left|
\frac{\pi}{14},-\frac{\pi}{14}
\right\rangle_{\text{bare}}
\\
&+0.07926
\left|
\frac{3\pi}{14},-\frac{3\pi}{14}
\right\rangle_{\text{bare}}
+0.17889
\left|
\frac{5\pi}{14},-\frac{5\pi}{14}
\right\rangle_{\text{bare}}
\\
&+0.66359
\left|
\frac{\pi}{2},-\frac{\pi}{2}
\right\rangle_{\text{bare}}
-0.44547
\left|
\frac{9\pi}{14},-\frac{9\pi}{14}
\right\rangle_{\text{bare}}
\\
&-0.12848
\left|
\frac{11\pi}{14},-\frac{11\pi}{14}
\right\rangle_{\text{bare}}
-0.03464
\left|
\frac{13\pi}{14},-\frac{13\pi}{14}
\right\rangle_{\text{bare}},
\end{aligned}
\end{equation}
The observable of interest is the projection of the undressed scattering state onto the undressed bound state, given by
\begin{equation}
\begin{aligned}
\braket{e | e^{-iHt}| \frac{\pi}{2}, -\frac{\pi}{2}} = -0.3766e^{-iE_{-} t} + 0.3663e^{-iE_{+} t} + O(\delta < 10^{-1}) \implies \\
\left|\braket{e | e^{-iHt} | \frac{\pi}{2}, -\frac{\pi}{2}}\right|^{2} \approx 0.55 \sin^{2} \left(\frac{E_{+} - E_{-}}{2} t \right)
\end{aligned}
\end{equation}
where we identify the avoided crossing gap $\Delta_{\text{AC}}$ as $|E_{+} - E_{-}| \approx 1.699$.

\subsection{Circuit Optimization and Compression for Hardware}
\begin{table}[htbp]
    \centering
    \begin{tabular}{lccc}
        \toprule
        \textbf{Compilation of Trotterized $U(\tau)$} &
        \boldmath$\mathbf{N}_{\mathbf{RZZ}}$\textbf{ scaling} &
        \textbf{Depth scaling} &
        \textbf{Fidelity} \\
        \midrule
        Unoptimized        & $\sim 150n$          & $\sim 40n$            & $100\%$ \\
        Local Optimization & $\sim 98n$           & $\sim 26n$            & $100\%$ \\
        AQC compressed     & Effective $\sim 24.5n$ & Effective $\sim 6.5n$ & $ > 93\%$  \\
        \bottomrule
    \end{tabular}
    \caption{Resource scaling for different compilation strategies as a
    function of Trotter steps $n$. ``Fidelity'' here is against the unoptimized compilation. The compression generally provided a
    reduction factor of $\times 4$ by compressing the circuit $U(\tau)$ to the same number of layers and gates at $\lceil \tau  / 4 \rceil$ .}
    \label{tab:compilation_scaling}
\end{table}
\begin{figure}[H]
\centering
\includegraphics[width=0.70\textwidth, height=6cm]{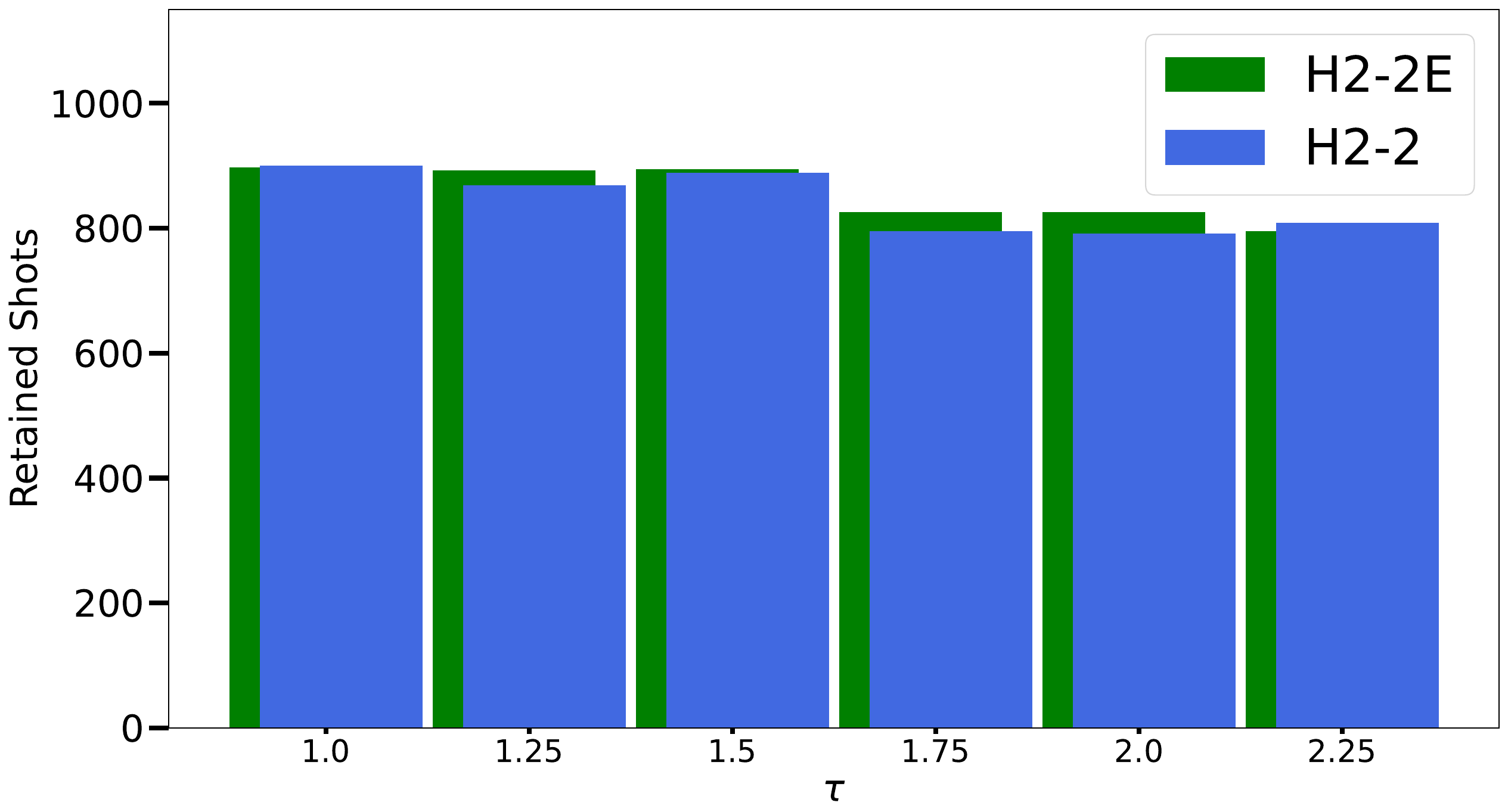}
\caption{Retained shots in the defined physical subspace in 1000-shot simulations.}
\label{fig:retained_shots}
\end{figure}
The Hamiltonian evolution circuits shown in section \ref{evolution_circuits} were optimized and compressed in order to run on Quantinuum \texttt{H2} without the need for significant error mitigation. This was achieved in two stages. First, the two-qubit gate count of the local 4-qubit unitary with fixed Trotter step size, $U(\text{d} \tau)$, was reduced through variational circuit
compilation. Second, the resulting circuits were compressed with AQC tensor-network optimization as a function of evolution time $\tau$.

For the $RZZ$ reduction, the hopping and interaction evolution terms were combined and variationally compiled into a fixed-parametrized quantum circuit. Optimization was done to maintain unit fidelity for the unitary within the physical plaquette subspace and between the physical and unphysical subspaces. A parametrized circuit ansatz with 6 $RZZ$ gates was found to reproduce the target evolution for the hopping and interaction coefficients considered. 

After this local optimization, we applied AQC tensor-network compression \cite{Robertson2025AQC} to the Trotterized unitary $U(\tau)$, with the goal to maximize the fidelity given a target depth at a smaller $\tau$. The target depth was chosen heuristically so as to maintain fidelity over 90\%. The resulting $RZZ$ scaling and circuit-depth are shown in Table \ref{tab:compilation_scaling}. 

All circuits on \texttt{H2-2} and \texttt{H2-2E} were run with 1000 shots with Quantinuum's custom dynamical decoupling \cite{QuantinuumH2DD}. The resulting shot data was split into physical and unphysical spaces, with the physical space being defined as the space of two-particle $+$ states and single-particle $e$ states. Any observables measured were first projected onto this physical space. Fig. \ref{fig:retained_shots} shows the retained shots for both the hardware and emulator simulations. 

\end{document}